\documentclass[preprint,11pt]{aastex}
\newcommand{\K}{\,\mbox{K}}
\newcommand{\erg}{\,\mbox{erg}}
\newcommand{\cm}{\,\mbox{cm}}
\newcommand{\s}{\,\mbox{s}}
\newcommand{\yr}{\,\mbox{yr}}
\newcommand{\Gyr}{\,\mbox{Gyr}}
\newcommand{\Mpc}{\,\mbox{Mpc}}
\newcommand{\mJy}{\,\mbox{mJy}}
\newcommand{\kpc}{\,\mbox{kpc}}
\newcommand{\kms}{\,\mbox{km}\,\mbox{s}^{-1}}
\newcommand{\msun}{\,M_{\sun}}

\newcommand{\amin}{\,\mbox{arcmin}}

\newcommand{\sfrunit}{\msun \yr^{-1}}

\newcommand{\Hz}{\,\mbox{Hz}}
\newcommand{\tsim}{\sim\!}
\newcommand{\ea}{et al.~}
\newcommand{\sfr}{{\it SFR}}
\newcommand{\sfrthick}{{\it SFR}_{\it thick}}

\begin{document}
\title{Sub-mm Galaxies in Cosmological Simulations}

\author{Mark A. Fardal, Neal Katz}
\affil{Astronomy Department, University of Massachusetts, Amherst, MA 01003}
\author{David H. Weinberg}
\affil{Astronomy Department, Ohio State University, Columbus, OH 43210}
\author{Romeel Dav\'e}
\affil{Steward Observatory, University of Arizona, Tucson, AZ 85721}
\author{Lars Hernquist}
\affil{Department of Astronomy, Harvard University, Cambridge, MA 02138}

\begin{abstract}
  We study the predicted sub-millimeter emission from massive galaxies in
  a $\Lambda$-dominated cold dark matter universe, using hydrodynamic
  cosmological simulations.  Assuming that most of the emission from
  newly formed stars is absorbed and reradiated in the rest-frame 
  far-infrared, we calculate the number of galaxies that would be detected 
  in sub-mm surveys conducted with SCUBA.  The predicted number counts are 
  strongly dependent on the assumed dust temperature and emissivity law.
  With plausible choices for parameters of the far-IR spectral energy 
  distribution (e.g., $T=35\K$, $\beta=1.0$), the simulation predictions 
  reproduce the observed number counts above $\sim 1\mJy$.  The sources have
  a broad redshift distribution with median $z \approx 2$, in reasonable
  agreement with current observational constraints.  However, the predicted
  count distribution may be too steep at the faint end, and the fraction of
  low redshift objects may be larger than observed.

  In this physical model of the sub-mm galaxy population, the objects detected
  in existing surveys consist mainly of massive galaxies (several $M_\ast$)
  forming stars fairly steadily over timescales $\sim 10^8-10^9$ years, at
  moderate rates $\sim 100\sfrunit$.  The typical descendants of these sub-mm 
  sources are even more massive galaxies, with old stellar populations,
  found primarily in dense environments.  While the resolution of our 
  simulations is not sufficient to determine galaxy morphologies, these 
  properties support
  the proposed identification of sub-mm sources with massive ellipticals
  in the process of formation.  The most robust and distinctive prediction of
  this model, stemming directly from the long timescale and correspondingly
  moderate rate of star formation, is that the far-IR SEDs of SCUBA sources
  have a relatively high 850~\micron\ luminosity for a given bolometric 
  luminosity.  Other model predictions can be tested with future studies of
  the redshift distribution, rest-frame UV/optical properties, and angular
  and redshift-space clustering of sub-mm galaxies.

\end{abstract}


\newcommand{\mapfig}{
\begin{figure*}
\plotone{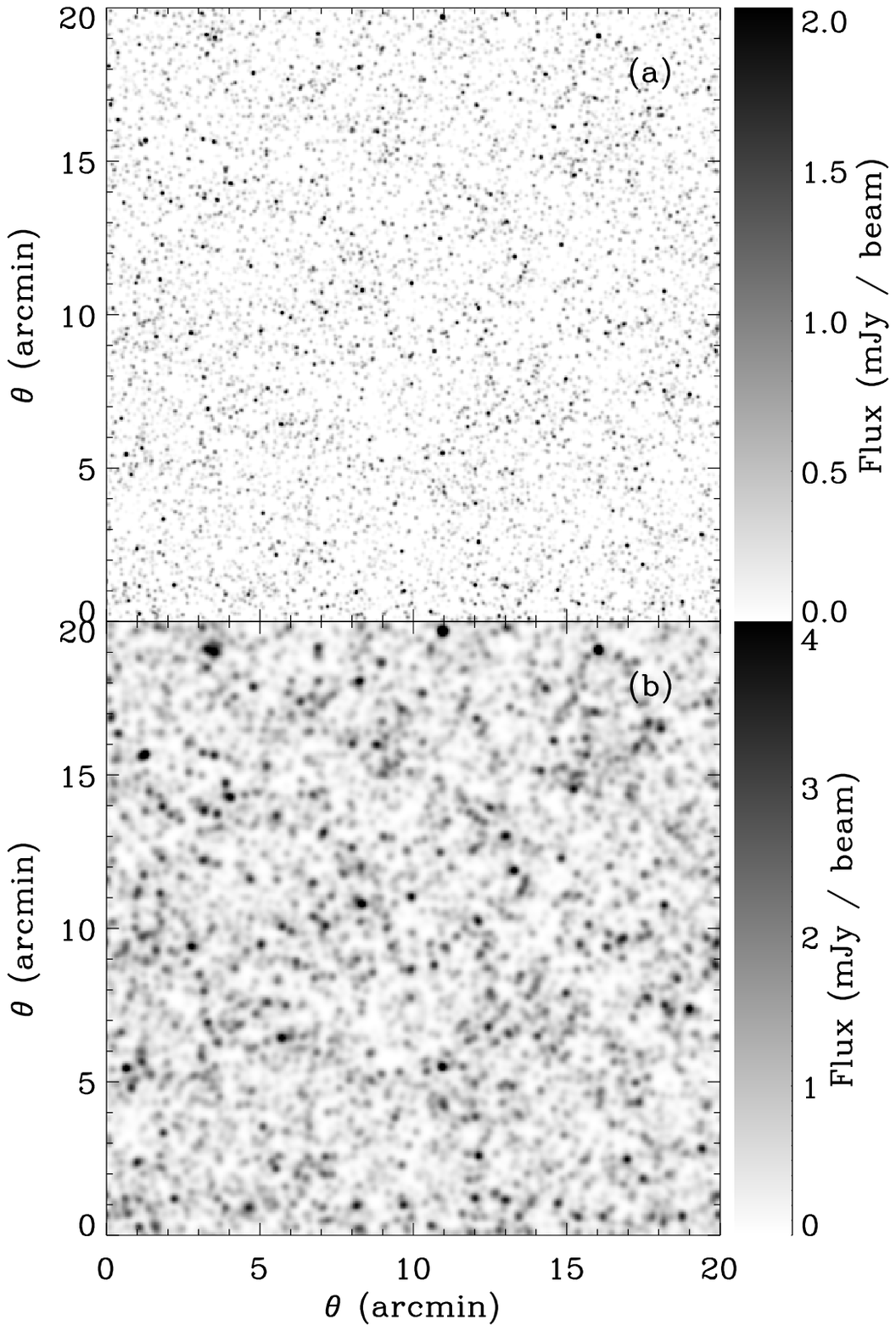}
\caption{
\label{fig.maps}
Simulated maps of the millimeter sky.  These are based on the 
L50/144 simulation, using our fiducial SED of $T = 35 \K$, $\beta = 1.0$,
and are generated by the randomization procedure discussed in the text.  
(a) This map is ``measured'' at a wavelength of 1.2~mm with a
resolution of 6 arcsec, appropriate to the Large Millimeter Telescope
(LMT) currently under construction.  The map is normalized such that a
pixel centered on a source with a given flux corresponds to the gray
scale level in the color bar.  
(b) 
This map uses a wavelength of
850~\micron\ and assumes a resolution of 15 arcsec appropriate to
SCUBA on JCMT.  Although SCUBA maps are usually quite noisy, no noise
is included in this or the previous map.
}
\end{figure*}
}

\newcommand{\countsfig}{
\begin{figure*}
\plotone{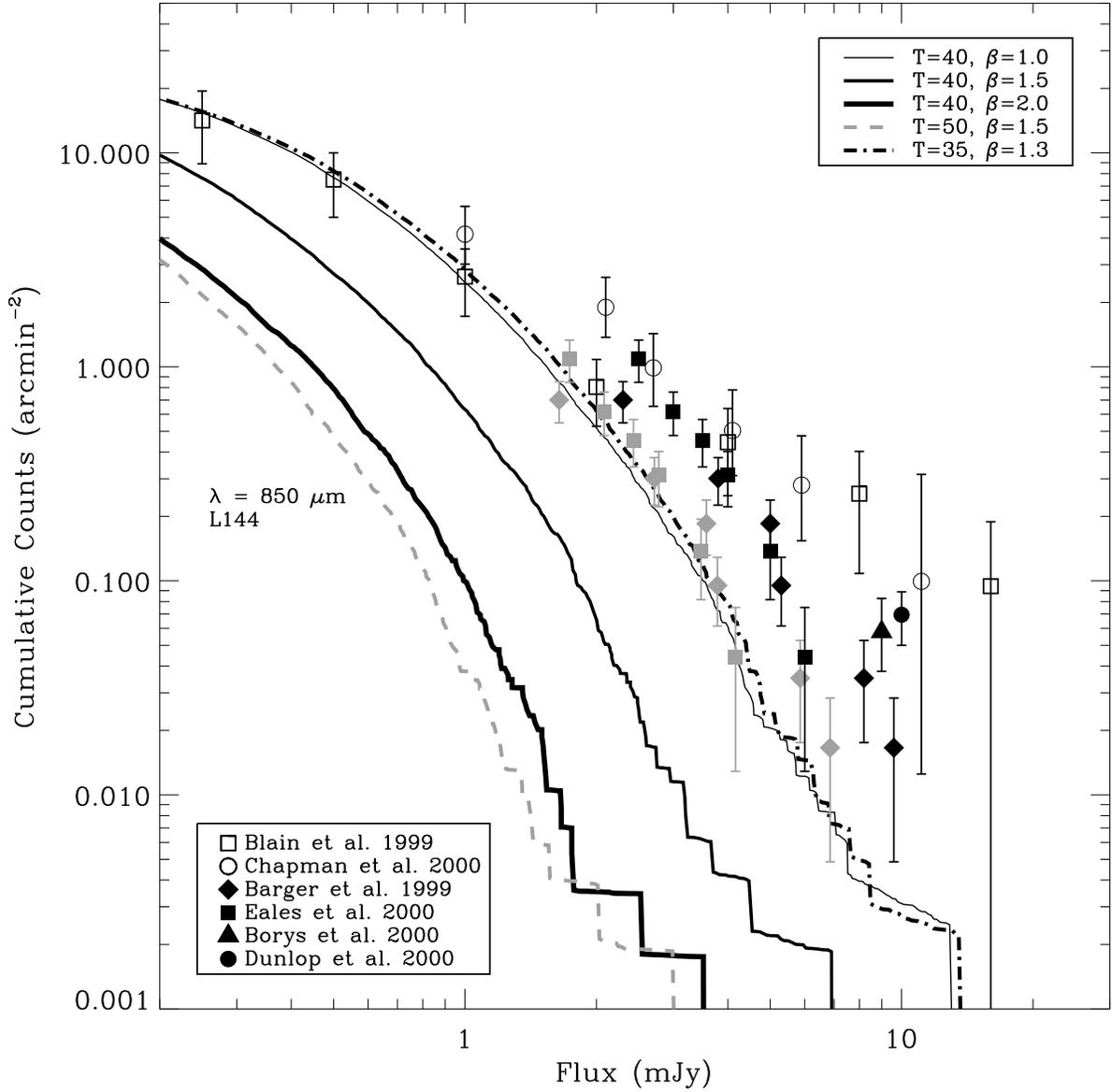}
\caption{
\label{fig.144counts}
Cumulative counts as a function of flux at 850~\micron, for
some single-temperature models.  Only the galaxies from the L50/144 simulation 
are
included here.  The SED parameters of the models are marked in the legend.
The various observational points are discussed in detail in the text;
open points come from lensed cluster fields, filled points come from
blank-field surveys, and grey points incorporate an approximate
(and somewhat controversial) correction for observational biases.
For the models listed in descending order in the figure legend,
the integrated backgrounds at 850~\micron\ are 
12.2, 6.2, 3.0, 2.7, and $12.9 \mJy \amin^{-2}$.}
\end{figure*}
}

\newcommand{\highfluxcountsfig}{
\begin{figure*}
\plotone{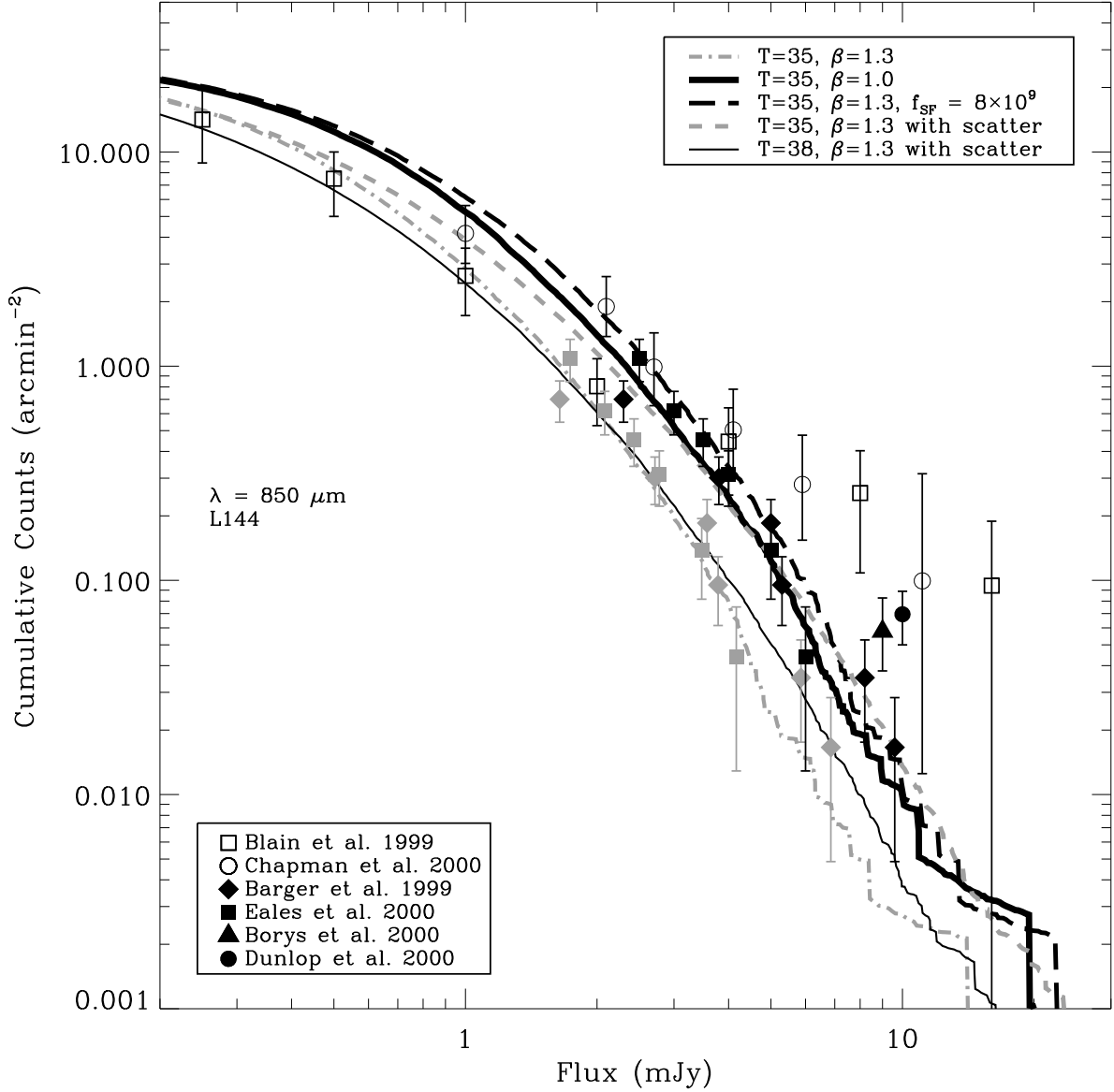}
\caption{
\label{fig.highfluxcounts}
Similar to Figure~\ref{fig.144counts}, but the predicted counts use modified
emission models that raise the counts at high fluxes as explained
in the text.
For the models listed in descending order in the figure legend,
the integrated backgrounds at 850~\micron\ are 
12.9, 18.7, 20.7, 15.2, and $11.3 \mJy \amin^{-2}$.  }
\end{figure*}
}

\newcommand{\medrescountsfig}{
\begin{figure*}
\plotone{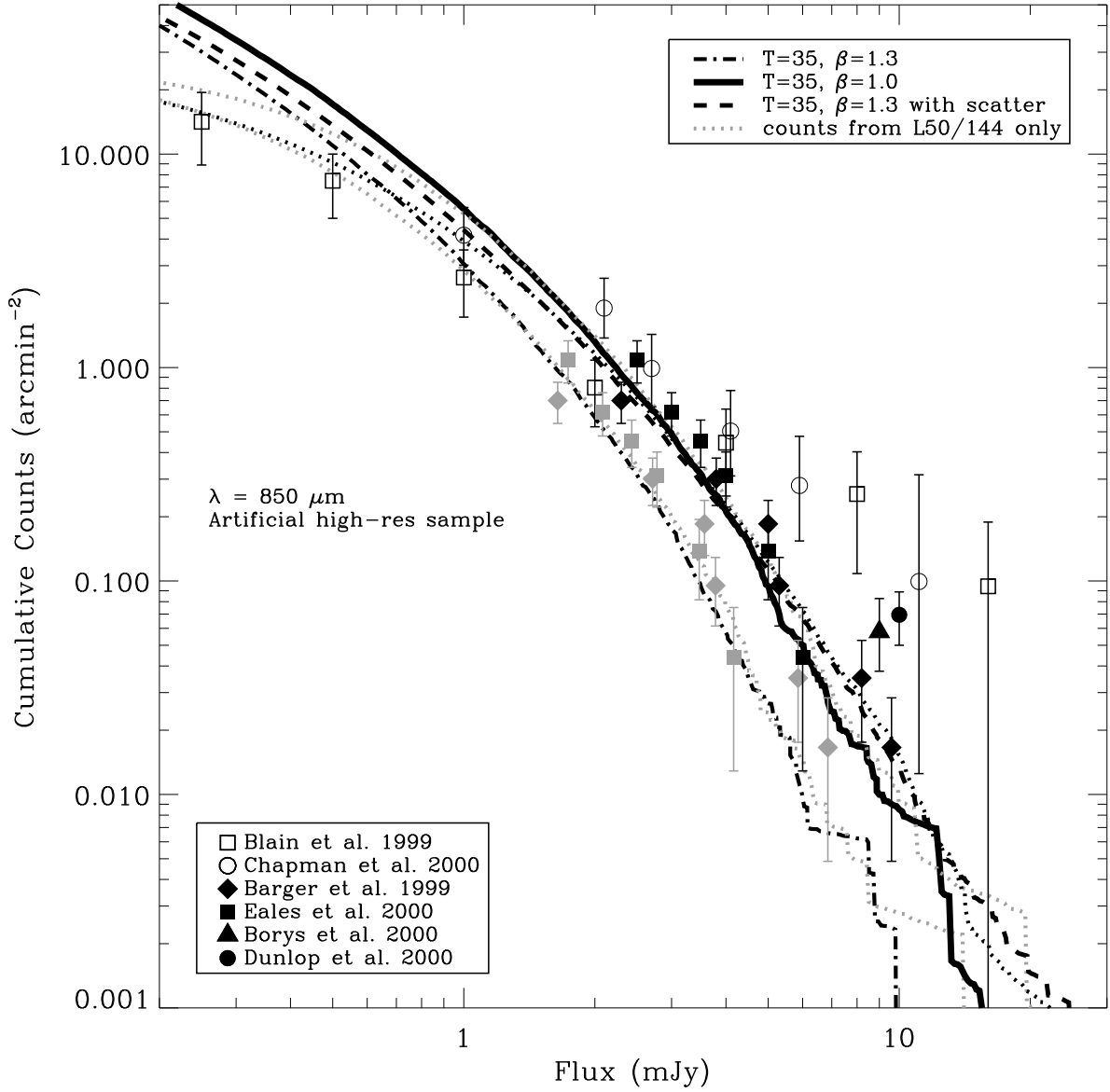}
\caption{
\label{fig.medrescounts}
Similar to Figure~\ref{fig.highfluxcounts}, but using results of the
artificial high-resolution sample.  Two of the models from the
previous figure are omitted here for clarity.  The dotted lines show
the results of the L50/144 simulation for comparison.  They use the
same set of emission models, and match onto the corresponding
high-resolution curves at about 1~mJy.  For the models listed in
descending order in the figure legend, the integrated backgrounds at
850~\micron\ from the high-resolution sample are 27.4, 37.3, and $31.3
\mJy \amin^{-2}$.
}
\end{figure*}
}

\newcommand{\zdistfig}{
\begin{figure*}
\plotone{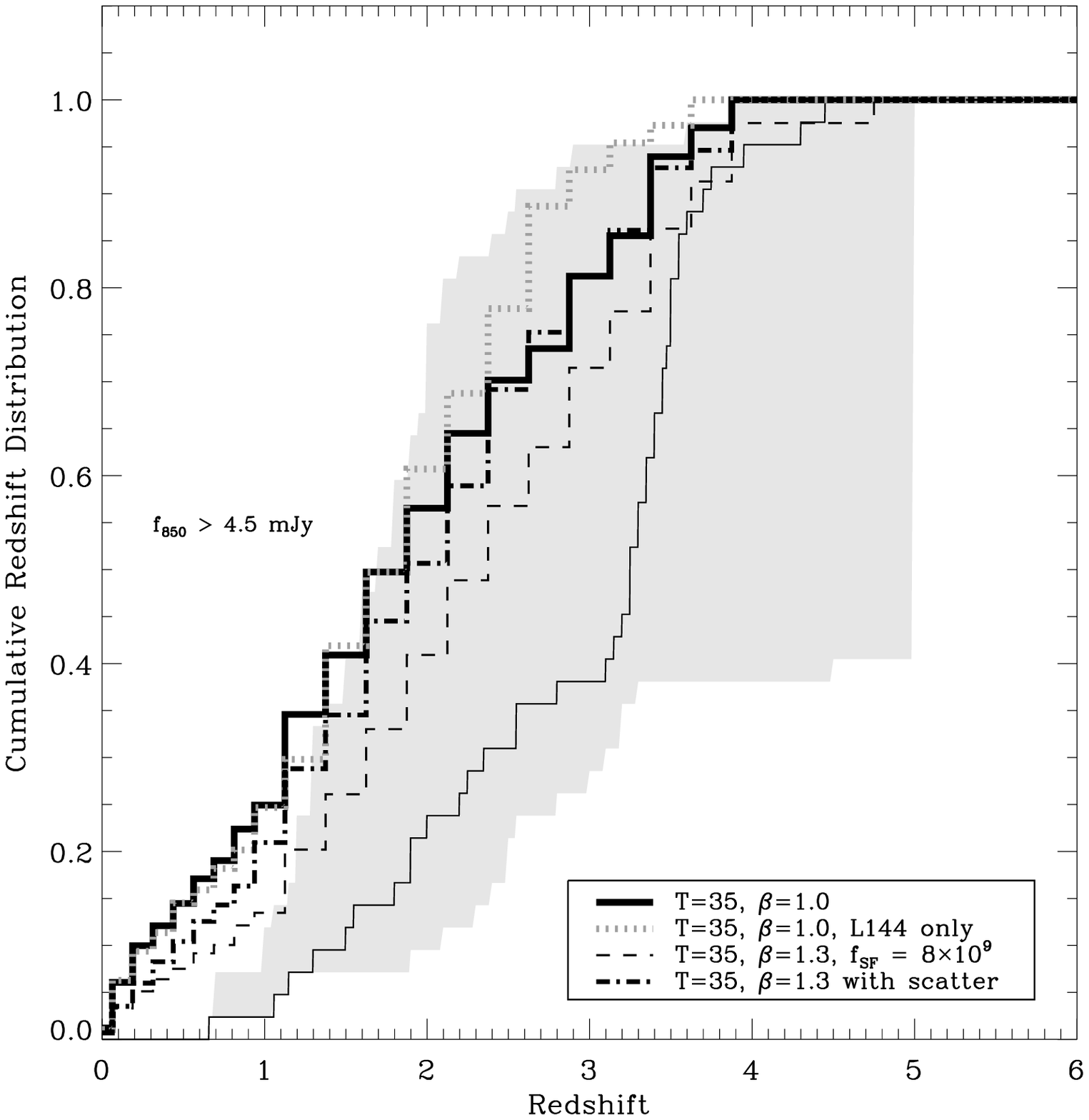}
\caption{
\label{fig.zdist}
The cumulative redshift distribution of the 850~\micron\ sources.  The
gray shaded region and the light solid curve bisecting it represent the
observational limits and mean estimates, respectively, from the combined
results of \citet{barger00,smail00zdist} and \citet{eales00}.  The
curves marked in the legend are based on our simulation, using various
emission models and assuming a sample threshold of $4.5 \mJy$.  The
stair-step appearance in the simulation lines
results from the use of discrete outputs.
The steps in the observational results are from individual sources.}
\end{figure*}
}

\newcommand{\massfuncfig}{
\begin{figure*}
\plotone{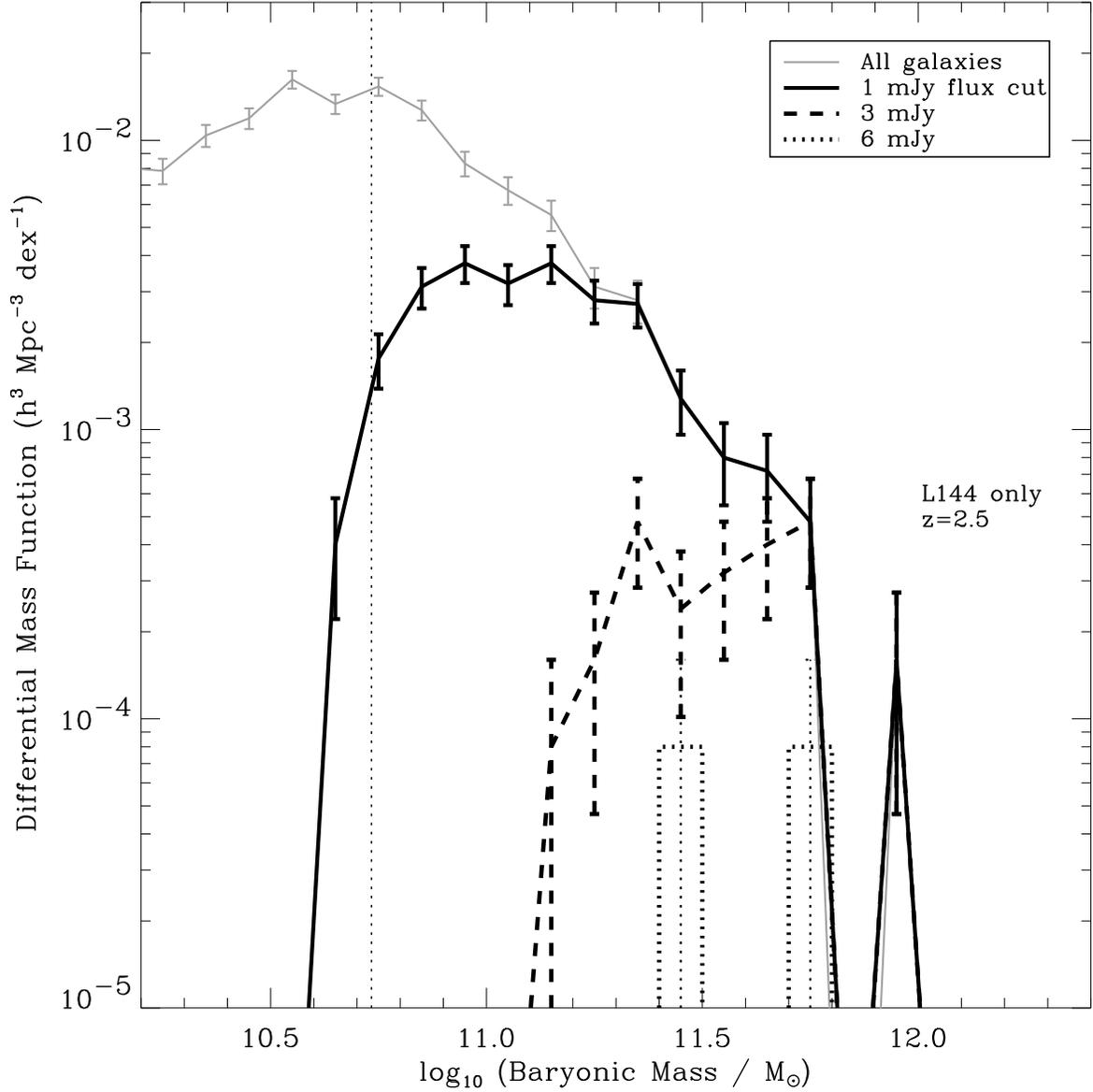}
\caption{
\label{fig.massfunc}
The distribution of baryonic mass 
in cold gas ($T < 3 \times 10^4 \K$) and stars
for the simulated galaxies.  The plot uses
the L50/144 simulation and assumes the fiducial SED model ($T = 35 \K$, $\beta =
1.0$).  Sub-samples selected by sub-mm flux are shown by the curves
marked in the legend.  Error bars assume Poisson statistics.}
\end{figure*}
}

\newcommand{\timedistfig}{
\begin{figure*}
\plotone{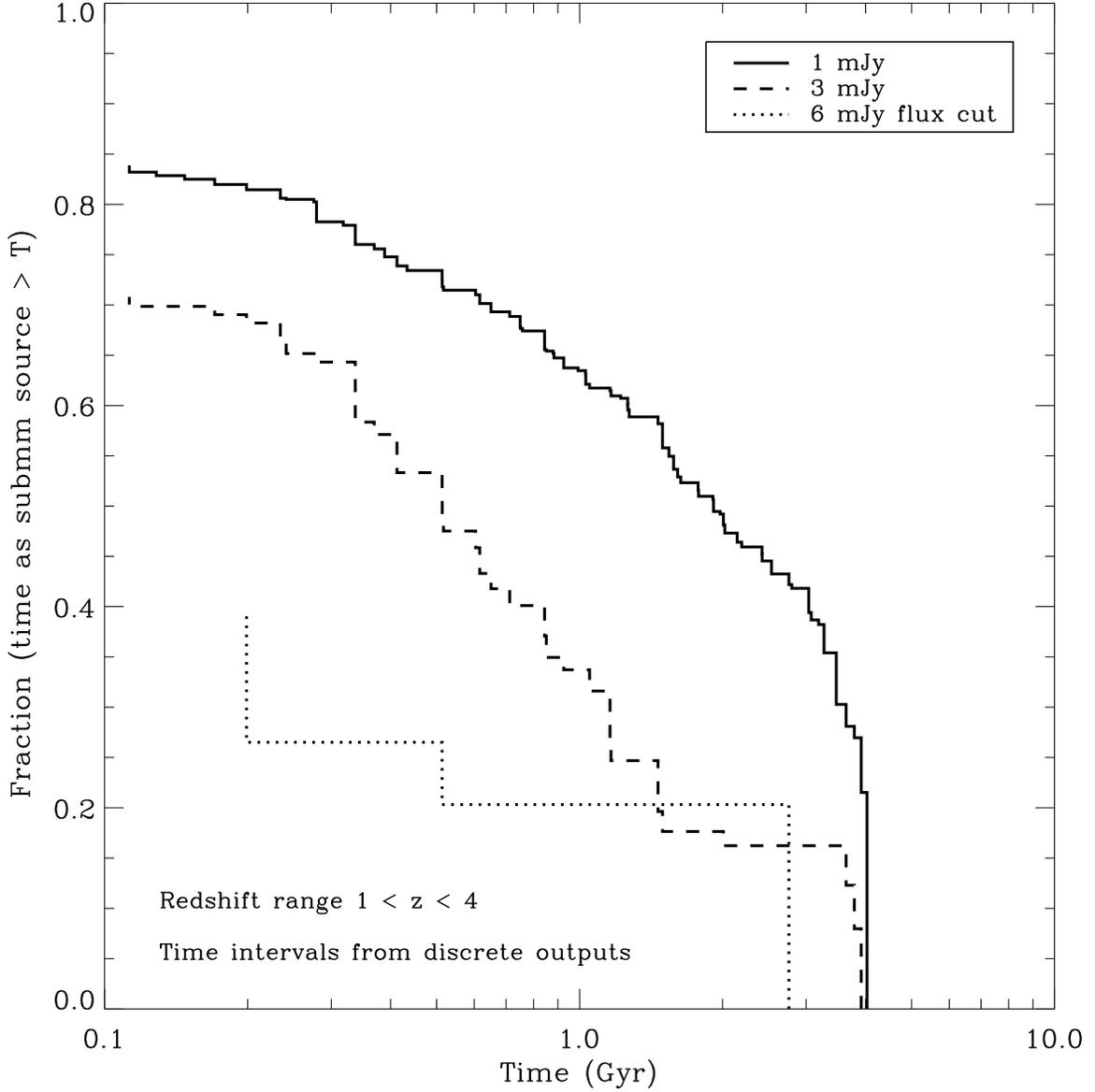}
\caption{
\label{fig.timedist}
Cumulative distribution of the lifespan of a sub-mm source, over
redshifts $1 < z < 4$ in the L50/144 simulation.  The distribution is
effectively weighted by the time a galaxy stays a sub-mm source, since
this is the way it would appear in an observational sample.  The
curves show results for different flux-limited samples.  
The spacing of our outputs increases from 0.1 Gyr at $z=4$ to 0.7 Gyr
at $z=1$, and we assign a lifespan $<0.1$ Gyr to any source that
does not appear in consecutive outputs.  With higher time resolution,
the true distribution would thus be higher than the plotted curves
below 0.7 Gyr.
}
\end{figure*}
}

\newcommand{\agefig}{
\begin{figure*}
\plotone{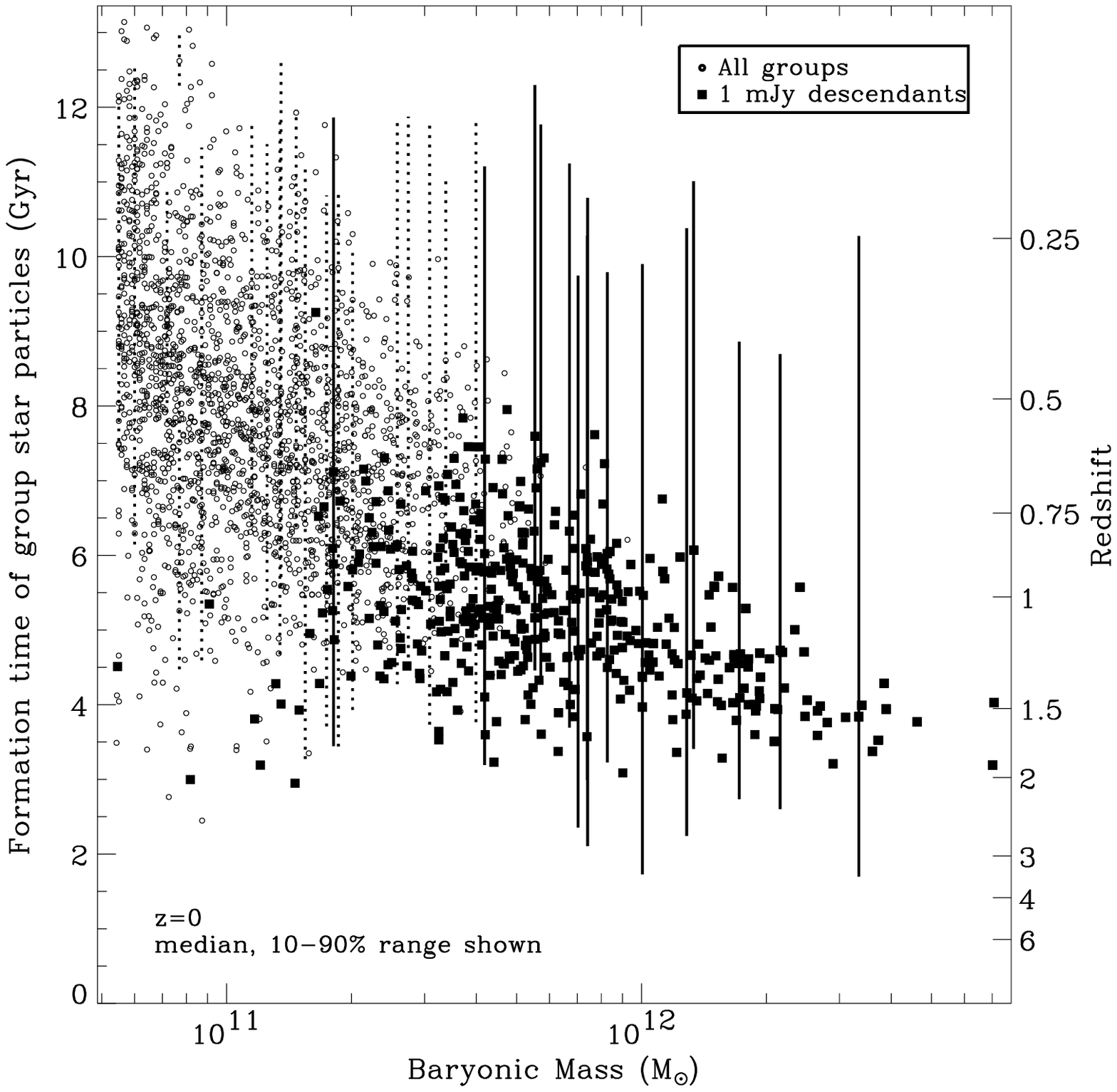}
\caption{
\label{fig.age_z0}
Median formation time of stars in the simulated galaxies versus their
baryonic mass in cold gas and stars.  
This plot uses only the L50/144 simulation.
More recent formation is at the top of the figure.  For randomly
selected galaxies, we have also shown the range where the middle 80\%
of the stars were formed.  Filled squares and solid lines show
descendants of the galaxies that were brighter than 1 mJy at
850~\micron\ at some redshift in the range $1<z<4$, using the fiducial SED
model.  Open circles and dotted lines show the remaining galaxies.  }
\end{figure*}
}

\section{INTRODUCTION}
The observation of galaxies in the sub-millimeter has become a crucial
area of astronomy since the installation of the SCUBA 
camera on the JCMT (the field is reviewed by \citealt{blain99review}).
Because emission from dust in galaxies rises sharply up to a peak at
$\tsim 100$ \micron, sub-mm observations preferentially detect
high-redshift galaxies.  While it was known from optical/near-IR
observations that the cosmic star formation rate increases steeply
with redshift up to at least $z \approx 2$ \citep{madau98},
measurement of the extragalactic background with FIRAS
\citep{fixsen98} suggested that still more of the high-redshift star
formation is hidden in the optical.  The first SCUBA surveys
\citep{smail97,hughes98,barger98} immediately showed that a large
fraction of this hidden star formation takes place in extremely
luminous systems, with star formation rates of $\gtrsim 100 \sfrunit$.

Many authors have conjectured that the objects dominating
the SCUBA surveys are elliptical
galaxies or spheroidal components of disk galaxies in the process of
formation.  The analogy to low-redshift galaxies that are similarly
bright in the far-infrared (FIR) suggests that they may be experiencing
bursts of star formation due to major mergers.  Although the surveys
are increasing in depth and quantity, the low resolution, small
fields, and large noise of the SCUBA maps compared to optical surveys
leads to many systematic uncertainties that are still poorly quantified.
Optical
counterparts of the SCUBA sources are still difficult to identify, and
there are very few detections in the FIR at wavelengths other than 
850~\micron.  As a result, vigorous debates continue about the redshift
distribution and star formation rates of the sources, the fraction of
their energy supplied by AGN and star formation, their connection to
other populations such as the Lyman break galaxies, and the
contribution of these sources to the cosmic star formation rate
\citep{lilly99,blain99model,adelberger00,eales00}.

Comparison of these sub-millimeter observations to current theories of
galaxy formation is clearly an important task.  Modeling of the SCUBA
population has so far been done in a phenomenological manner.  For
example, \citet{blain99model} and \citet{blain99semimodel} 
showed there must be rapid luminosity
evolution in the infrared galaxy luminosity function
to reproduce the observed counts and sub-mm background.
\citet{guiderdoni98} found that matching the counts within their semi-analytic
model of galaxy formation required the {\it ad hoc} addition of a
component of starbursting systems whose numbers increase rapidly
with redshift.  Artificial catalogs of sub-mm sources have also been
constructed using Poisson-distributed sources \citep{eales00} or
large-scale dark matter simulations \citep{hughes00}, to aid in
measurement of observational systematics.  While useful, such catalogs
do not bring one any closer to a physical understanding of these
sources.

In this paper, we use numerical N-body/hydrodynamical simulations to
study the sub-mm galaxy population.  This paper follows several others
devoted to the galaxy population in our simulations
\citep{katz99,weinberg99,davelbg99,weinberg00}.  
These papers showed that the
optically detected Lyman-break galaxies at $z \approx 3$ are a natural
consequence of CDM models, assuming reasonable amounts of dust
extinction.  They focused on the dependence of the galaxy abundance on
the assumed cosmology and the clustering of these sources.
This paper concentrates on the correspondence between the galaxies in
our simulation and the bright sub-mm sources detectable with SCUBA.
We restrict the study to a single cosmology, the currently favored
$\Lambda$-dominated cold dark matter
(LCDM) model.  We rely primarily on a single simulation of a large
volume but only moderate mass resolution, which is well suited to
a statistical study of the bright, rare sources.  We defer a detailed
discussion of fainter sources, the FIR background, and the relation
between the sub-mm and UV/optical populations to a future paper
using higher resolution, smaller volume simulations that are currently
underway \citep{fardal01}.

We describe our numerical simulations in \S~\ref{sec.galaxies},
where we also discuss the resulting galaxy population and examine the
effects of numerical resolution.  In \S~\ref{sec.emission}, we
describe our recipe for calculating the sub-mm fluxes.  Our method
ignores the detailed physics of dust absorption and emission, opting
instead for simple, parameterized models motivated by observations.  In
\S~\ref{sec.observables}, we constrain these parameters by requiring
the simulated galaxy population to match the counts at $\tsim 3 \mJy$.
We compare the resulting flux and redshift distributions to
observations, arguing that the sub-mm population can be reproduced in our
simulations for plausible assumptions about dust emission.
In \S~\ref{sec.physical}, we examine the physical
properties of the simulated galaxies that correspond to SCUBA sources.
Section \ref{sec.discussion} discusses our confrontation of the simulations
with observations and presents our conclusions.

\section{SIMULATING THE GALAXY POPULATION}
\label{sec.galaxies}
Our simulations are performed with the N-body/hydro code Parallel
TreeSPH.  This code follows both the gas and the dark matter with
discrete particles, using spline kernel interpolation to smooth the
gas quantities over a compact set of particles.  The code is described
elsewhere in detail \citep{hk89,katz96,dave97code}.  Simulations with this
code have been successful in reproducing the observed clustering
pattern and evolutionary trends of galaxies \citep{katz99,weinberg00},
as well as phenomena such as the Lyman-$\alpha$ forest
\citep{hernquist96,dave99}, and the abundance of higher-column
density systems, including Lyman-limit and damped Lyman-$\alpha$
absorbers (e.g. Gardner et al. 1997a,b; 2000).

Star formation in this simulation is determined heuristically using
local physical properties of the gas.  For star formation to take
place within a gas particle, the gas must be cold ($T < 3 \times 10^4
\K$), dense ($\rho_H > 0.1 \cm^{-3}$ and overdensity 
$\rho_g/{\bar\rho_g}>55.7$), 
converging, and Jeans unstable.
The star formation timescale is determined by the local cooling and
dynamical timescales, and because the star formation rate is an 
increasing function of gas density, the star formation rate in a simulated
galaxy is governed mainly by the rate at which gas condenses and
cools onto the central object (see \citealt{katz96} for details of
the algorithm and \citealt{weinberg00} for further discussion).
We identify galaxies in the
simulations at the sites of local baryonic density maxima, using the program
SKID\footnote{http://www-hpcc.astro.washington.edu/TSEGA/tools/skid.html}.
SKID finds galaxies by first determining the smoothed density field,
then moving particles upward along the gradient of the density field.
We define the galaxy to be the set of particles that aggregate at a
particular density peak after removing particles that do not satisfy a
negative energy binding criterion.  To be included in a galaxy, gas
particles must also have a temperature $T<3\times 10^4\K$ and and a
density $\rho > 10^3\Omega_b$, i.e.  we consider only the cold gas and
stars, the material that comprises the bulk of the matter in the
central, visible regions of galaxies.  Adding up the star formation
rates for the individual particles in the galaxy gives the total star
formation rate for that galaxy.  Emission from active galactic nuclei
or quasars is not included in the simulation.

Properties of the simulated galaxies are described in
\citet{weinberg99} and \citet{weinberg00}.  Galaxies in the
simulations form with a 2-phase gas structure: a central glob of gas
at $\tsim 10^4 \K$ that forms stars, surrounded by hotter gas at
$\tsim 10^6 \K$.  Galaxies with luminosities comparable to the Lyman
break galaxies form in abundance even at $z > 5$.  The clustering of
these high-redshift galaxies is highly biased relative to the mass, in
accordance with observations \citep{katz99}.  The star formation rates
in the simulated galaxies are relatively steady on timescales of
200~Myr.

Comparing simulations of different resolutions and volumes, we find
that there is a regime in which the galaxy properties are robust to
changing these numerical parameters.  The criterion for the low-mass
end of this regime is that the galaxy should have $\tsim 64$ cool gas or star
particles.  At the high-mass end, the volume should be large enough to
allow halos of a given size to form in significant numbers.  This
latter criterion is less easy to quantify (as it is affected not only by
Poisson statistics but also by the exclusion of long-wavelength modes
in finite boxes), but it can be seen empirically as an upper 
cutoff in the mass function that depends on the box size.  The star
formation rates of the galaxies are strongly correlated with the
galaxy masses, especially at high redshift.  Hence, the star formation
rate also has a characteristic resolution limit, though it is not as
sharply defined as the mass limit.  Because of the restricted range
of halo masses, any one simulation represents
only a portion of the total star formation.

We refer to the main simulation used in this paper as the L50/144
simulation.  This simulation is tuned to study the evolution of large 
($L \gtrsim L_\ast / 4$)
galaxies down to $z=0$. It has $144^3$ particles in each of the 
gas and dark matter components, and models a 
periodic cube that is $50 h^{-1}$ comoving $\Mpc$ on a
side.  This gives us
a baryonic mass resolution of $5.4 \times 10^{10} \msun$ (based on the
64-particle limit above), and a spatial resolution of
$7.0 h^{-1}$ comoving $\kpc$ (equivalent Plummer softening).  
We assume a $\Lambda$
dominated cold dark matter cosmological model with $\Omega_m = 0.4$,
$\Omega_\Lambda = 0.6$, $h \equiv H_0 / (100 \kms \Mpc^{-1}) = 0.65$
and a primeval spectral index $n=0.93$.  With the tensor mode
contribution, normalizing to COBE using CMBFAST
\citep{seljak96,zaldarriaga98} implies a normalization $\sigma_8=0.8$,
which provides a good match to cluster abundances \citep{white93}.  We
use the \citet{hu96} formulation of the transfer function.  We adopt a
baryonic density $\Omega_b = 0.02 \, h^{-2}$ consistent with the
deuterium abundance in high-redshift Lyman-limit systems
\citep{burles97,burles98}, and the opacity of the high-redshift
Lyman-$\alpha$ forest (e.g. Rauch et al. 1997).

For describing the faintest sub-mm sources and for computing the FIR
background, the L50/144 simulation has insufficient resolution.  We are
conducting a suite of simulations with the same cosmology as above, but with
different resolutions and box sizes to test the sensitivity of our
results to these parameters.  In this paper, we supplement the
results of the L50/144 simulation with those of two additional simulations.  
L11/64 uses
$64^3$ particles in an $11.1 h^{-1} \Mpc$ volume evolved to $z=0$;
this simulation has a mass resolution of $6.8 \times 10^9 \msun$ and a
spatial resolution of $3.5 h^{-1}$ comoving $\kpc$.  The other simulation, 
L11/128, uses $128^3$ particles in an $11.1 h^{-1} \kpc$ volume
evolved to $z=3$; this simulation has a mass resolution of $8.5 \times
10^8 \msun$ and a spatial resolution of $1.75 h^{-1}$ comoving $\Mpc$.
When combined, these simulations give consistent results for the
number density of galaxies with different star formation rates at high
redshift (see
\citealt{weinberg99}).  

Although we will focus mainly on the bright sources that should be adequately
represented in the L50/144 simulation, for some purposes we want to 
consider the contribution of lower mass systems.  To do so, we fit
the star formation rate (SFR) distribution across the L144/50, L64/11,
and (at $z>3$) L128/11 simulations, using a \citet{schechter76} function,
which turns out to be a reasonable fit:
\begin{eqnarray}
  \frac{dn}{d(\mbox{\it SFR})} 
         & = & \frac{\phi_{\ast}}{\mbox{\it SFR}_\ast}
          (\mbox{\it SFR} / \mbox{\it SFR}_\ast)^{\alpha_s}
          e^{-\mbox{\it SFR} / \mbox{\it SFR}_\ast} \nonumber \; , \\
  \mbox{\it SFR}_\ast & = & \mbox{\it SFR}_\ast^0 
     (1+z)^{\delta_s + \gamma_s \ln(1+z)} \; ,
\label{eq.sfrstar}
\end{eqnarray}
with $\phi_{\ast} = 6.49 \times 10^{-3} \, h^3 \Mpc^{-3} \mbox{dex}^{-1}$,
$\alpha_s   = -1.96$,
$\mbox{\it SFR}_\ast^0 = 33.9 \sfrunit$,
$\delta_s = 2.63$, 
and $\gamma_s = -1.04$.  This fit is in good agreement with the L50/144
simulation in its valid range of star formation rates.  
We choose a lower cutoff on the SFR to correspond to the completeness
limit of the L11/64 simulation, which roughly equates to 
${\it SFR}_{\it min} = 1.4 \, (1+z)^{1.6} \msun \yr^{-1}$.
We then generate an
artificial high-resolution ``galaxy'' population from these fits.
Naturally, this artificial sample can be used only as an indicator of
the results expected from a higher-resolution, large-volume
simulation.  It also cannot be used to study the
spatial distribution or merger history of the galaxies.
Further discussion of our fitting methods and results with
a fuller set of simulations will be presented in \citet{fardal01}.

The star formation rates in the simulations are the foundation of this
paper, so we review here what we already know or suspect about the
validity of these rates.  For the cosmological model considered here,
the simulations reproduce the observed density of Lyman-break galaxies
given a reasonable amount of extinction \citep{weinberg00, davelbg99}.  The
cosmic star formation rate rises from $z=0$ back to $z \approx 2$--3 
and then falls
off, though we are still testing the sensitivity of the high-redshift decline
to resolution \citep{weinberg99}.  Comparison to low-redshift
observations in the UV \citep{sullivan00} and FIR
\citep{saunders90} suggests that the simulated star formation rates at
low redshift may be too high by a factor $\tsim 2$, although the
observed functions and conversions from luminosity to SFR
are subject to much uncertainty.
The total amount of stars formed by $z=0$ may also be too high by a
factor $\tsim 3$ \citep{weinberg99}.  Tests in \citet{katz96} show
that the star formation rates are robust to changes in the algorithm
parameters, being determined by gas supply rather than these
parameters, although an entirely different algorithm could no doubt
change the rates.

In sum, there are various hints that the overall amount of star formation
in the simulations may be too high, although the size of the discrepency
depends heavily upon the assumed IMF.
This discrepancy could indicate
observational underestimates of the true star formation or stellar density,
numerical errors in the simulations, problems in the cosmological model
(e.g., incorrect values of $\Omega_b$ or $\sigma_8$), incorrect assumption 
about the form or constancy of the stellar initial mass function, 
or an incorrect treatment of the ``microphysics'' of star formation and
feedback.  A major goal of comparing simulations to a diverse set
of observations is to gain guidance on this issue.

The SCUBA sources may be unusual galaxies, and even simulations that
are adequate for explaining typical galaxies may be inadequate for
studying rarer objects.  This concern is particularly acute because
the rarity of the SCUBA sources forces us to use a large simulation
with low resolution.  The true star formation rates may depend on
details of galaxy structure that are unresolved in our
simulation.  For example, in simulated mergers involving two
large spirals \citep{mihos94a, mihos96} or between spirals and 
less-massive companions \citep{mihos94b, mihos95}, the
peak star formation rate depends
strongly on the presence or absence of a bulge.  In our simulations the
low spatial resolution and large particle noise generally prevent the
formation of galactic disks.  However, if star formation in
the SCUBA sources does not involve nuclear starbursts, then
the resolution of the simulation might not be crucial.  At any
rate, it is not clear {\it a priori} whether or not we can accurately
obtain the luminosities of the sub-mm galaxies, and the reader should
keep these sources of uncertainty in mind.

\section{EMISSION MODEL}
\label{sec.emission}
Ideally, to model the emission from our galaxies we would take the
distribution of stars and gas, compute the spatially
dependent dust opacity by tracking the gas temperature and
metallicity, and solve the radiative transfer problem
through the distributed opacity.  
However, our simulations are far from adequate for this purpose:
they do not resolve the internal structure of galaxies on the
necessary scale (that of individual star-forming clouds), and they
do not track the metallicity and dust content of the gas.
We will therefore pursue a much simpler approach, working from the
global star formation rate in each galaxy to a sub-mm flux using
empirically motivated recipes.

The first step is to calculate the bolometric luminosity of each
galaxy.  In actively star-forming galaxies, the bolometric luminosity
is dominated by young stars and is therefore roughly proportional to the star
formation rate.  We assume a Miller-Scalo \citep{miller79} initial
mass function (IMF) extending from 0.1 to $100 \msun$, consistent with
the star formation treatment in our code.  Using the code STARBURST99
\citep{leitherer99}, and assuming a burst length of $10^8 \yr$ and a
metallicity of $Z = 0.02$, we derive a conversion factor of
\begin{equation}
f_{SF} \equiv L_{bol} / ({\it SFR}) = 5.0 \times 10^9 L_\sun \msun^{-1} \yr \; .
\label{eq.sfrconv}
\end{equation}

We emphasize here that the calibration of the luminosity in the parameter
$f_{SF}$ is quite poorly known.  For one thing, there is a dependence
on the burst age due to the buildup of intermediate-mass stars.
\citet{blain99model} use
$f_{SF} = 2.2 \times 10^9 L_\sun \msun^{-1} \yr$ for the same IMF,
presumably assuming a burst length of $10^7 \yr$.  (The Miller-Scalo
IMF is very sensitive to burst length below $3 \times 10^7 \yr$,
because of its steep slope above $10 \msun$.)  There is also a
strong dependence on the shape of the IMF, particularly at low
masses, since the mass is concentrated in low or medium-mass stars but
the luminosity is concentrated in high-mass stars.  A Salpeter IMF
with the same mass range, burst age, and metallicity used
in equation~(\ref{eq.sfrconv}) gives
$6.5 \times 10^9 L_\sun \msun^{-1} \yr$.  The modification of the
Salpeter IMF at the low-mass end suggested by 
\citet{kroupa00} gives only 73\% as much mass, or 
$f_{SF} = 8.9 \times 10^9 L_\sun \msun^{-1} \yr$.  Variations in the
IMF have been much discussed, usually with the implication that more
rapidly star-forming galaxies may have top-heavy IMFs and hence be
more luminous (e.g., \citealt{doane93}).  Variation in the metallicity
can also affect the luminosity by at least several tens of percent.

We adopt the value in equation~(\ref{eq.sfrconv}) as our standard
conversion factor.  However, the true value of $f_{SF}$
parameter could easily differ by a factor of two, and it could
even vary systematically from one type of galaxy to another.

The next step is to determine the fraction of this luminosity that is
absorbed and re-radiated by dust.  
The particular extinction model we choose is motivated by
several observational facts.  
Extinction estimates for Lyman-break
galaxies at $z \approx 3$, based on the UV spectral index, suggest
large (mean of 5--8) and variable amounts of extinction
\citep{adelberger00}.
We also require high average extinctions for our simulations to match 
the rest UV luminosity function for these galaxies
\citep{weinberg00}.
Furthermore, the bright
SCUBA sources tend to be extremely faint in direct optical/UV light
\citep{chapman00radiosel}.
All of these facts suggest that star-forming galaxies at high redshift
are highly opaque, and that the most luminous ones are
the most heavily absorbed, as is the case at low redshift.

The specific model we use assumes that the
internal extinction of a galaxy depends upon its star formation rate.  
Let $f_{\it UV}$ and $f_{\it FIR}$ be the fractions of the
stellar energy that escape in the UV-optical continuum or are absorbed
by dust, respectively.
We assume that the average ratio FIR/UV ratio in a galaxy with
star formation rate {\it SFR} is
$f_{\it FIR} / f_{\it UV} = {\it SFR} / \sfrthick$,
where $\sfrthick=4\sfrunit$ is the {\it SFR} at which 50\% of
the bolometric luminosity is absorbed.
We also introduce a random scatter in this ratio of 0.5
dex per galaxy.  This model makes
the UV emission uncorrelated with the FIR emission at high bolometric
luminosity, and it results in a Schechter-like cutoff 
in the UV luminosity function for Lyman break galaxies.
With these opacities, the SFR cutoff in the high-resolution sample 
corresponds roughly to the population
of Lyman-break galaxies seen at $z \approx 3$ for $R_{AB} \lesssim 27$.

The extinction model here is not unique, and we will examine 
this issue in more detail in \citet{fardal01}.
However, the essential point of the
model for our purposes is that galaxies with large star formation
rates are typically highly obscured; this is certainly true for the
majority of the SCUBA sources.  Our results in this paper thus
depend only weakly on the details of the extinction, except perhaps
for the estimates of the integrated background.
Conversely, the predictions for the bright end of the UV/optical
luminosity function are much more sensitive to extinction assumptions.

The next step is to assign an FIR spectral energy distribution (SED)
to these galaxies.  We adopt the usual emission model with just three
parameters: a dust temperature $T$; an emissivity index $\beta$, such
that the emissivity at low frequencies is $f_\nu \propto \nu^{2 +
  \beta}$; and a short-wavelength slope $\alpha$, such that $f_\nu$
falls no faster than $\nu^\alpha$.  In luminous high-redshift
galaxies, these parameters are of order $T = 40 \K$, $\beta
= 1.5$, and $\alpha \approx -1.7$ (e.g.\ \citealt{blain99review}).
Increasing $T$, $\beta$, or $\alpha$ decreases the flux at 
850~\micron\ for galaxies at $z \lesssim 5$.  These parameters provide
reasonable, though not exact, fits to the SEDs of galaxies at low and high
redshift \citep{dunne00,ivison98}.  We frame our discussion of SEDs
in terms of these parameters, not because they are truly fundamental
parameters of the galactic emission (which in a more realistic model
would come from dust at a variety of temperatures)
but because they provide a simple and commonly used description.

The dependence of our results on the parameters $T$ and $\beta$ will
be examined below.  The short-wavelength slope $\alpha$ is unimportant
for determining the shape of the spectrum near 850~\micron, but it 
results in an overall shift in the normalization.  Hence we keep it
fixed at a value $\alpha = -3.0$, comparable to that seen in Arp 220
\citep{hughes00}.  For this choice, the quasi-thermal portion of the
spectrum is shifted downwards by about 15\%.  The short-wavelength
behavior of the spectrum may be the most variable part in the SEDs of
low-redshift, FIR-bright galaxies (e.g. Figure 1 in \citealt{hughes00}).
Values of $\alpha = -2.2$ or $-1.7$, as assumed by
\citet{blain99model} or \citet{blain99review}, reduce the quasi-thermal
spectrum by factors of 1.4 or 1.9 respectively (assuming $T = 40 \K$,
$\beta = 1.5$).

The following fitting formula may be convenient for comparison with
other work.  With our choices for $\alpha$ and $f_{SF}$, the 
850~\micron\ flux from a completely opaque galaxy at $z=2$ is
approximately given by
\begin{equation}
f_{850} = (0.44 \mJy) \exp[- 3.7 t (1 - 0.30 t)] \, f_{\it FIR} \, 
(\mbox{\it SFR}/100 \sfrunit), 
\end{equation}
where 
$t \equiv \ln(T / 40 \K) + 0.38 (\beta-1.5)$.
This formula is accurate
to within 20\% over the range $25 \K < T < 60 \K$, $0.3 < \beta < 2.5$.  
Similar fits can be
constructed at other redshifts, with slightly different coefficients.  
The effects of changing $T$ and $\beta$ on the flux are nearly degenerate.

For simplicity, most of our models will use a single SED
for all galaxies. However, there seems to be substantial scatter in
the observed SEDs at low redshift \citep{dunne00,adelberger00}.
Using a SCUBA survey of the IR-luminous galaxies in the local universe
complemented with IRAS data, \citet{dunne00} fitted $T$ and $\beta$
for each galaxy, finding they were drawn from populations with
means and standard deviations of $T = 35.6
\pm 4.9 \K$ and $\beta = 1.3 \pm 0.2$.  Below we will test the effect
of this scatter on our predictions.

Our empirical emission models are heavily based on the properties of
IR-luminous galaxies at low redshift.  In high-redshift galaxies, the
amount of dust will probably be lower due to lower metallicity, while
the FIR luminosity will typically be higher; these effects can be
expected to raise the temperature.  On the other hand, the FIR
spectrum is also sensitive to the dust properties through the value of
$\beta$, in a way that is not quantitatively understood.  It is thus
difficult to say whether the conclusions drawn at low redshift will
also apply at higher redshift.

\section{COMPARISON TO SUB-MM OBSERVABLES}
\label{sec.observables}
To compute predictions of sub-mm counts, we take the
list of star formation rates by particle group (``galaxy'') 
at each output redshift.  The 27
outputs used from the L50/144 simulation are at redshifts $z=0$, 0.125, 0.25...
1.0, 1.25, 1.5..., 4, 4.5, 5... 6, and 7.  Conceptually, we take the
simulated volume and replicate it over shells centered on these
redshifts.  Each galaxy enters the observed distribution in an amount
proportional to the volume of its redshift shell.  The galaxies in any
output are partly descended from those in previous outputs, of course,
so the different outputs are correlated.  We use the parametric models
in \S~\ref{sec.emission} to construct catalogs of observed flux, and
thus distributions of the sources in flux and redshift.  We construct
similar catalogs based on the analytic fit to the high-resolution
results (eq.~\ref{eq.sfrstar}).

We also construct simulated sub-mm maps based on our L50/144
simulation. Figure~\ref{fig.maps}a shows a map at 1.2mm using the
angular resolution of the large millimeter telescope (LMT) currently
under construction.  Figure~\ref{fig.maps}b is at 850~\micron\ and
uses the angular resolution of SCUBA on JCMT.  Neither map includes
any observational or instrumental noise.  These maps use a
randomization procedure much like that employed by \citet{croft00}.
Since the sources all have angular extents much smaller than the $15
\, \arcsec$ SCUBA beam, we treat each galaxy as a point source.  The
simulation box is replicated many times in comoving space along the
line of sight, using the redshift output closest to the required
redshift.  Each box is given a random offset in the $x$, $y$, and $z$
directions, using the periodic nature of the simulation.  The box is
given a random parity inversion along each axis, and an axis is chosen
randomly to lie along the line of sight.  The box is then spun by a
random amount around the line of sight.  The resulting field size is
20 arcmin on a side, since we include outputs to $z=7$ and lose some
area due to the spinning part of the randomization algorithm.  As
discussed below, these maps are incomplete with respect to sources
below 1 mJy compared to a higher-resolution simulation.

Figures~\ref{fig.maps} give a visual sense
of how the current SCUBA observations relate to the future observations
that will be possible with the LMT.  We also use these simulated maps
in our analysis of clustering of sub-mm sources discussed in 
\S\ref{sec.physical} below.

\mapfig

\subsection{Distribution in flux}

Using the catalogs of fluxes from each galaxy in the simulation, we
calculate the counts on the sky as a function of flux at 850~\micron.
The counts that result from using only the L50/144 simulation, and
single-temperature models, are shown in Figure~\ref{fig.144counts}.
This plot demonstrates the large effect that the uncertainty in the
SED can have on the predicted counts.  The emission models used span
most of the range usually considered for the sub-mm sources.  Raising
either $T$ or $\beta$ depresses the calculated fluxes by reducing
the 850~\micron\ luminosity for a given bolometric luminosity.
Also, $T$ and
$\beta$ are nearly degenerate in their effect on the counts, as one can
see, for example, by comparing
the curves for $T=40 \K$, $\beta = 1.0$ and $T=35
\K$, $\beta=1.3$.  As discussed above, the short-wavelength slope
$\alpha$ could shift the curves leftward by tens of percent.  
The uncertainty in $f_{SF}$ could also shift the curves horizontally
by as much as a factor of two (in either direction).

Some of the results from SCUBA surveys of galaxies are presented on
the plot, with their source marked in the legend.  The many different
surveys and the large scatter they exhibit deserve some explanation.
The solid symbols show the results of blank-field surveys
(\citealt{barger99,eales00,borys01,dunlop01}).  The lensing cluster
surveys (\citealt{blain99obs2,chapman00survey}) have two advantages
stemming from the magnifying effect of the cluster: they can go deeper
in flux, and source confusion is reduced.
However, the necessary correction for the
cluster lensing amplification introduces some risk of systematic
error.

Since this is a cumulative plot, the data points are correlated in any
given survey.  Also, the error bars generally account for Poisson
errors only and thus may be underestimated.

The raw counts are strongly affected by observational biases, because
of the poor spatial resolution and S/N currently available.
\citet{eales00} argue that the combined effect of noise and confusion
amplifies the measured source fluxes by about a factor of 1.4 on average
in their survey, independent of flux.  We have thus shown in gray
their survey, and the similar survey of \citet{barger99}, corrected
down by this factor.  However, this argument is controversial,
with other workers disagreeing on the constancy \citep{blain00} or
even the sign \citep{hughes00} of the effect.  It is also unclear whether
the cluster surveys, or the higher flux limit, blank-field surveys
\citep{borys01,dunlop01}, are affected by confusion; if not, the
disagreement in the measured counts is even greater
at the high-flux end.

The sub-mm observations are still clearly in their early stages, and
the present results have a large scatter that is not all due to
random statistical errors.  If we take our results with $T \approx 35 \K$ and
$\beta=1.3$, roughly the mean parameters found in the local SCUBA
survey \citep{dunne00}, we appear to have a reasonable fit to the
lower envelope of the observations defined in part by the
confusion-corrected points.  This model would produce an integrated
850~\micron\ background of
$12.9 \pm 4.7 \mJy \amin^{-2}$,
which agrees with the background measurement from FIRAS,
$(12.2 \pm 4.7) \mJy \amin^{-2}$ \citep{fixsen98}.  The background
measurements from DIRBE \citep{hauser98} exceed those from FIRAS
by about 40\% at shorter wavelengths, although the two measurements are
formally consistent, so the true background may be slightly higher.  This
SED is at the ``cold'' end of the parameters usually considered for
sub-mm sources, an issue we return to in
\S~\ref{sec.observables.sed}.

\countsfig

While this model is in agreement with the lowest of the observational
points, it is hard to escape the impression in
Figure~\ref{fig.144counts} that the true counts are somewhat higher.
(Whether this is so depends largely on the validity of the confusion
correction introduced by \citealt{eales00}.)  
It is interesting to see what kinds of models are
required for the simulations to match the raw blank-field or
the cluster survey counts instead.  In
Figure~\ref{fig.highfluxcounts} we show three attempts to accomplish
this match.  The first model simply adopts yet smaller SED parameters
of $T=35 \K$, $\beta=1.0$.  
The second model uses a larger $f_{SF} = 8
\times 10^9 L_\sun \msun^{-1} \yr$, with $T=35 \K$ and $\beta=1.3$,
and produces similar results to the first.  In both cases the simulated
counts may be steeper than the observations.

The third and fourth altered models allow
for random scatter in the galaxy SEDs, as
discussed in \S~\ref{sec.emission}.  In a real galaxy the 
factors determining dust emission are
surely not ``random'', but this model reflects our ignorance of what
causes the variation in the FIR SEDs of local galaxies.  We find that the
counts become significantly shallower in this case, improving the
agreement with the blank field surveys. 
A mean temperature of $T = 35 \K$ gives a good fit to the uncorrected
counts, while a temperature of $T = 38 \K$ provides a better fit to the
confusion-corrected points.


Because of its physical plausibility and its reasonable success
in matching the uncorrected blank-field counts,
we choose the $T = 35 \K$, $\beta = 1.0$ model as our fiducial model
for the rest of the paper (e.g., for Figures~1 and 7--9).
However, to zeroth order, choosing a different $T$ and $\beta$
combination affects the fluxes from all galaxies in the same way,
although there are first order differences depending on redshift.
Hence, any such combination that fits the uncorrected counts will
contain essentially the same objects at any given flux or counts
threshold, and thus imply almost the same physical properties.  Also,
if one accepts the shift in the fluxes suggested by \citet{eales00},
the preferred SED parameters change as discussed above; but the
implied physical properties for a given flux-limited sample should be
almost the same if one simply lowers the flux thresholds by a factor 
of 1.4.

\highfluxcountsfig

So far, the calculated counts use only the L50/144 simulation.  However,
the typical star formation rate at a given baryonic mass is higher at high
redshift (when gas accretion rates are higher), so the minimum
star formation rate that is resolved in any given
simulation rises towards higher $z$.  With our preferred flux models, this
resolution threshold
begins to affect the counts at the 2 mJy level by $z \sim 3$ in the
L50/144 simulation.  To examine the effect of resolution, we use the artificial
high-resolution sample discussed in \S\ref{sec.galaxies}.  Since this sample is
generated from an analytic fit rather than an actual simulation, it
should be viewed as merely illustrative of the effect of higher numerical
resolution. However, its properties are well constrained by the
behavior in the low and high resolution simulations.

Figure~\ref{fig.medrescounts} illustrates the effect of this extended
high-resolution sample on the simulated counts.  For each model, we
show the results obtained both with the extended sample and with the L50/144
simulation alone.  
Extending the sample makes a significant difference below 1~mJy.
The models that best agree with the observed counts above 1 mJy no
longer agree below that level, and their cumulative surface brightness 
now exceeds the FIRAS
background.  Since the background values obtained using only
the L50/144 simulation are acceptable,  a cutoff in the opacity at $\tsim 60
\sfrunit$ could produce agreement, but this cutoff value is likely too
high.  

The predicted excess of faint sources seen in
Figure~\ref{fig.medrescounts} may be a genuine problem for our model.
However, at low luminosities and $z<3$, our analytic fit to the SFR
function is constrained mostly by the small L11/64 simulation; hence
uncertainty in the fit could be the cause of the discrepancy.  The 
background excess could also be influenced by our simplistic
extinction model.  In this paper we are mainly interested in the
objects above 1--2~mJy, so we postpone a detailed examination of this
issue to a future paper \citep{fardal01}.

The counts in the L50/144 simulation may be incomplete at the bright
end as well, due to the finite volume and consequent exclusion of
long-wavelength modes.  This probably occurs at a counts level of
$\tsim 0.01 \arcmin^{-2}$, or a flux level of $\tsim 10 \mJy$ with our
fiducial model, though it is difficult to confirm this without running
a simulation in a yet larger volume.

\medrescountsfig

With the best-fitting emission models here, the star formation rates
implied for the SCUBA galaxies are large, though smaller than some
estimates.  For example, at $z=2.5$, a galaxy with $\sfr=100 \sfrunit$
has a flux of $1.2 \mJy$ at 850~\micron, assuming our fiducial model
of $T=35 \K$ and $\beta=1.0$.  For comparison, the flux would be only
$0.16 \mJy$ using $T=40 \K$, $\beta=1.5$, $\alpha=-2.2$, and 
$f_{SF} = 2.2 \times 10^9 L_\sun \msun^{-1} \yr$ as in
\citet{blain99model}.

\subsection{Distribution in redshift}
\label{sec:zdist}
\zdistfig

The cumulative redshift distribution of the sources 
is shown in Figure~\ref{fig.zdist}.  
Here we assume a flux threshold $f(850\, \micron) > 4.5 \mJy$,
comparable to the observational samples discussed below.
The distribution is actually bimodal, as shown by the 
subtle minimum of the slope at $z \sim 0.5$.  
The broad high-redshift portion extends to $z
\approx 4$.  Using our fiducial SED model and the artificial
high-resolution sample, the median redshift for the 4.5 mJy sample is
2.0; the mean and standard deviation are 2.0 and 1.3.  Using only the
L50/144 simulation lowers the redshifts, with a median of 1.8, mean of 1.8,
and standard deviation of 0.9; the shifts are small because of the
high sample threshold.  Using a higher $\beta$ and $f_{SF}$ as
indicated in the legend increases the median redshift to 2.4, with a mean of
2.3 and standard deviation of 1.2.  
If the \citet{eales00} confusion correction is applied to all fluxes
in these redshift surveys, an appropriate model as discussed above is
$T = 35 \K$, $\beta=1.3$, and our standard $f_{SF}$; but in this case
one should use a sample threshold of $\tsim 3.0 \mJy$.  Since lowering
the flux threshold and raising the fluxes from all the objects are
equivalent changes, the resulting distribution is nearly identical to
the curve with the non-standard $f_{SF}$.

The predicted redshift distribution is not strongly affected by our
choice of SED among the alternatives we prefer above.  The main impact
is to change the balance of the low-$z$ and high-$z$ components.  The
bimodal distribution arises in our simulation for the following
reason.  At high $z$, the flux from a source of a given luminosity is
nearly constant, but the flux rapidly increases as $z^{-2}$ for $z \ll
1$.  The mean luminosity of our sources brightens as a power-law in
$(1+z)$ at low $z$, so the characteristic flux reaches a minimum at $z
\sim 0.5$.  
The low-$z$ component is boosted by the low values of $\beta$
in our preferred models.
Changing the threshold flux also changes the balance of
the low-$z$ and high-$z$ components. Above 10 mJy we find that about
half of the sources are at $z < 0.5$ for our fiducial model. However,
at this flux level the counts in our simulations are very
incomplete at high $z$, due to finite-volume effects.  
For a threshold of 2 mJy, we find instead a higher median redshift 
of 2.3 for our fiducial model.

Observational constraints on the redshift distribution of the sub-mm
population are still fairly weak.  Few sources have spectroscopic
redshift measurements, so most of the redshifts are estimates based on the
radio-FIR correlation.  The derived redshift is dependent on the
assumed SED, and is roughly proportional to $T$.  In many cases the
radio emission is undetected, giving only a lower limit to the
redshift.  The redshift estimates and limits from
\citet{smail00zdist}, \citet{barger00}, and \citet{eales00} all assume
a set of SEDs to obtain the allowed redshift ranges.  We have combined
them here without regard to the slightly different SEDs employed.

Assuming a maximum redshift of $z_{\it max} = 5$ for all sources, we
have placed the observed sources at their lower redshift limit, their
upper redshift limit, or at the middle of the range (the limits are
taken from the observational papers and are $\tsim 2\sigma$).  The
extreme assumptions give the shaded region in Figure~\ref{fig.zdist},
while the middle assumption gives the solid line bisecting this
region.  Most of the difference between these curves is systematic
error due to the uncertainty in the SED.  With the lower values of $T$
and $\beta$ that we prefer above, the redshift distribution would be
close to the left hand boundary.  The mean of the estimated redshifts
(i.e., the middle curve, which assumes higher SED parameters than
preferred here) is 2.9,\footnote{Attempting to incorporate the fact
  that about half of these sources have only lower redshift limits,
  \citet{eales00} estimate the mean redshift to be $2.81 \pm 0.36$
  from the same surveys, not significantly different from our value.} 
and the median is 3.3.  The dispersion in the estimated redshifts of
the sources is 0.91.  Of course, some of this dispersion could be due
to scatter in the SED or the radio emission properties, or to
observational error.

Our redshift distribution is in fairly good agreement with these
results, with a similar mean (once the effect of the SED is accounted
for) and width.  We do appear to have an excess of
predicted sources at $z < 1$.
This excess could be related to the large amount of low redshift star
formation in our simulations.  We are currently investigating this
issue with a larger set of simulations \citep{fardal01}.
The brightest observed sources could include sources
that are primarily AGN, which we do not include in our modeling here;
their inclusion would probably boost the predicted median redshift.

\subsection{The SED of the sub-mm sources}
\label{sec.observables.sed}
The SED parameters of our preferred models are consistent with
low-redshift observations of FIR-bright galaxies \citep{dunne00}.
Somewhat higher temperatures have generally been preferred for SCUBA
sources in the literature, however, which raises the question of whether
our temperatures are too low to be plausible.  \citet{smail97}
preferred a temperature of $60 \K$ and described $T = 40 \K$ as ``very
cold''.  Other choices have included $T=47 \K$, $\beta=1.0$
\citep{barger98}; $T=50\K$, $\beta=1.5$ \citep{hughes98}; and $T=40\K$,
$\beta=1.5$ \citep{blain99review}.  Changing $f_{SF}$ by a reasonable
amount would bring our best-fit temperatures into this range, as shown
above.

As noted by \citet{eales00}, the argument that the majority of SCUBA
sources are ultraluminous infrared galaxies 
(ULIRGs) is somewhat circular.  If the temperatures are
assumed to be 40--60 K, typical of local ULIRGs, rather than 20 K,
typical of local normal spirals, one finds that the sources
have the bolometric luminosity of a ULIRG, but lower assumed temperatures
would imply lower luminosities.
Escaping the circularity requires redshifts and multiwavelength
observations of sub-mm sources without evidence of AGN activity.
There are only a few such sources at present
(see discussion in \citealt{blain99temp}).  These do seem
to have temperatures of roughly 40--50 K, though the limits are not
stringent.
In principle, an understanding of the dust mass implied by a given
SED combined with an upper limit on the metals in a galaxy's ISM
can give lower limits on the temperature.  The temperatures assumed
in our best-fit models do not appear to violate these limits.

Some direct evidence regarding the SEDs of SCUBA sources comes from the 
450~\micron\ measurements of \citet{eales00}, who derived a $3\sigma$
upper limit of
$\langle S_{450} \rangle / \langle S_{850} \rangle < 2.4$ for their
850~\micron-selected SCUBA sample.  Eales \ea argue that this limit
implies surprisingly low dust temperatures ($T \lesssim 25$, if
$\beta=2.0$) for the sub-mm sources.  However, we disagree with this
inference.  First, if we accept their argument that their 850~\micron\ 
fluxes are biased high by a factor 1.4 from observational confusion,
this limit increases to 3.4. \footnote{ \citet{blain99obs2} report
  detections of four galaxies at 450~\micron, but they do not estimate
  the mean flux ratio for their sample.  Since their derived surface
  density at 450~\micron\ and 10~mJy is equal to that at 850~\micron\ 
  and 3~mJy, we may loosely infer a flux ratio of $\langle S_{450}
  \rangle / \langle S_{850} \rangle = 3.3$, at least consistent with
  this limit.}  Second, \citet{eales00} assume a fixed redshift
to argue that the temperatures are low.  However, these redshifts are
ultimately themselves based on an assumed SED, because they are almost
entirely based on the radio-sub-mm flux ratio.  As discussed by
\citet{blain99temp}, this ratio depends essentially on the combination
$T / (1+z)$.  The 450/850~\micron\ flux ratio depends on the
same combination of parameters.  Hence assuming cool dust temperatures
alters the implied redshifts, and there is little net effect on the
450/850~\micron\ ratio; the two parameters are degenerate with
respect to these two flux ratios.  

However, the effects of $\beta$ and $(1+z)/T$ on the 450/850~\micron\ 
and radio/sub-mm ratios are not entirely degenerate; $\beta$
has a stronger influence on the former ratio than the latter.  Hence the
low 450 \micron\ fluxes seem to suggest low values of $\beta$.
Let us make a simple calculation, representing the entire sub-mm
population by a single ``typical'' source.
We calibrate the radio luminosity using the results for
low-redshift IRAS sources of \citet{yun01}.
Including their conversion factor of 1.5 between IRAS
luminosity and total dust luminosity $L_{\it bol}^{\it{}(dust)}$,
the implied FIR-radio correlation is
$S_{1.4 GHz} = 3.1 \times 10^{18} \erg \s^{-1} \Hz^{-1} 
( L_{\it bol}^{\it{}(dust)} / L_\sun )$.  
We also assume a radio continuum slope of $-0.8$.  We require that the
850 \micron\ / 1.4 GHz ratio be 100, typical of the sub-mm source
population, and the 450/850~\micron\ ratio be $< 3.4$.  Then we find
the limits $\beta \lesssim 1.0$ and $(1+z)/(T/ 40 K) \lesssim 2.8$.
Admittedly, this calculation is crude and the input parameters
are fairly uncertain.  However, it does seem to support the low values
of $\beta$ used in our preferred SEDs.

Unfortunately, further multiwavelength observations of the FIR
continuum alone will be just as inadequate in disentangling the temperature
and redshift.  The key to determining the source SEDs is to obtain the
source redshifts by some other means: from optical spectroscopy or
photometry, mid-IR spectral features, or molecular lines.  At this
point, we know very little about the correct choice of SED, which is
so important when comparing the simulations to observations.  All we can say is
that the observations may marginally favor the lower $\beta$ values
that we have suggested, and do not contradict the lower values of $T$.

To summarize our results in this section, there is a galaxy population
in our simulations that corresponds to the observed sub-mm sources, but
only if the dust temperatures are sufficiently low or the stellar
energy output is sufficiently high.  To zeroth order the agreement is
acceptable, although the counts may be somewhat too steep and there
may be too many low-redshift objects compared to current observations.
We now go on to describe the physical properties of these galaxies.

\section{PHYSICAL PROPERTIES OF SUB-MM GALAXIES}
\label{sec.physical}
In general, our sub-mm sources are quite massive galaxies, as shown in
Figure~\ref{fig.massfunc}.  In a flux-limited sample with 
$S_{850} > 1 \mJy$, the mass in cold gas and stars has a median of
$M_b \gtrsim 1.3 \times 10^{11} \msun$ at $z=2.5$ (assuming $T=35 \K$ and
$\beta=1.0$), and the median mass increases with increasing flux
threshold ($3.8 \times 10^{11} \msun$ for 3 mJy and $4.4 \times
10^{11} \msun$ for 6 mJy).  The distribution is quite broad, however.
If we choose instead an SED with $T = 35 \K$ and $\beta = 1.3$, the
star formation rate at a given flux increases by a factor of 1.4;
since the star formation rate is roughly proportional to mass, the
masses at a given flux threshold also increase approximately by this
factor.  The stellar mass distribution (not shown here)
is yet broader, with a median
stellar mass of $6.6 \times 10^{10} \msun$ at 1 mJy, again increasing
with flux.

We can define a timescale for star formation in these galaxies by
taking this median mass over the median star formation rate of $116
\sfrunit$, giving a typical timescale of $0.60 \Gyr$.  For reference,
the cosmic time is $2.4 \Gyr$ at $z=2.5$.  Thus the high star
formation rates of our simulated galaxies are not just due to
extremely short bursts.  This conclusion is also implied by the
correlation at high redshift between star formation rate and mass in
our simulations, particularly for the high-mass objects
\citep{weinberg00}.  If discrete bursts were the dominant
cause of the high luminosities, they would
tend to randomize the mass-luminosity correlation. We note that if the
emission model contains random scatter as discussed above, the
distributions of the flux-limited samples in Figure~\ref{fig.massfunc}
become broader, but the general trends remain.

\massfuncfig

Another way to characterize the size of our sub-mm galaxies is to find
their circular velocities $V_c$.  However, if the galaxies live in
groups or clusters, there could be an important distinction between the
circular velocity of the galaxy itself and that of the largest virialized
structure containing it.
To tackle this question, we first use the program
FOF\footnote{http://www-hpcc.astro.washington.edu/TSEGA/tools/fof.html}
to group dark matter particles into halos by the friends-of-friends
algorithm, using a linking length appropriate to the mean interparticle
spacing at the edge of a virialized halo \citep{kitayama96}.
We center a sphere on each
sub-mm galaxy and measure $V_c$
at a radius of 10 physical $h^{-1} \kpc$, the smallest reasonable 
radius given our spatial
resolution of $2 h^{-1} \kpc$ at this redshift. We also center the sphere on the
most bound particle in the FOF halo to which the sub-mm galaxy
belongs, and measure $V_c$ at the radius where the density matches
the virial density.

The circular velocities defined at $10h^{-1}\kpc$ and at the halo virial radius
are not
greatly different, at least at $z=2.5$.  The median $V_c$ of the 1, 3,
and 6 mJy samples are 328, 500, and $569 \kms$ for the galaxies
themselves at $10 h^{-1} \kpc$.  
They are 282, 403, and $495 \kms$ for
their surrounding virialized halos at the virial radius.  
It is not surprising that the
galaxy circular velocities somewhat exceed those of their surrounding
dark matter halos.  Each sub-mm galaxy tends to be the dominant 
baryonic object in its neighborhood, with a mass large enough to
increase the circular velocity at $10 h^{-1}\kpc$.
At these redshifts, large virialized
structures with velocity dispersions characteristic of clusters 
(and higher than those of their central galaxies) have
yet to form.  
The high circular velocities, substantially higher than those
of $L_*$ galaxies today, confirm that the sub-mm galaxies in
this simulation are massive objects residing in massive dark matter
halos.

We have computed the time evolution of the L50/144 simulation up to the
present day, so we can directly measure the lifetimes of the sub-mm
sources.  To do this, we define various 850~\micron\ flux-limited
samples of simulated galaxies, assuming $T = 35 \K$ and $\beta = 1.0$.
We then construct a merger tree of all of these galaxies down to
$z=0$, using the same methods as in \citet{murali01}.  Restricting our
study to the redshift interval $1 < z < 4$, which includes most of the
SCUBA sources (c.f.\ Figure~\ref{fig.zdist}), we find the distribution
of lifetimes for sub-mm sources as shown in
Figure~\ref{fig.timedist}.  Our time resolution is limited by the
spacing of our discrete simulation outputs, which varies from $0.1
\Gyr$ at $z=4$ to $0.7 \Gyr$ at $z=1$.  If a source does not repeat
(i.e., appear in more than one output),
we assume that it lasts for less than 0.1 Gyr.  
With higher time resolution, therefore, the time distribution
curves would be higher at lifespans $<0.7$ Gyr, but the fraction
of non-repeating sources is small, so this should not be a large effect.
Of course, we are
unable to tell whether the emission from a given galaxy turns off
briefly between outputs; but if our sub-mm sources were constantly
flickering on and off, this would only reduce the apparent lifetimes.
The mean age of a 3~mJy source is $\sim 500$ Myr, although a significant
fraction of our sources remain sub-mm sources for
more than 1 Gyr.  The contribution to the background is even more
heavily weighted towards the long-duration sources than is apparent in
Figure~\ref{fig.timedist}, as they tend to have larger luminosities.
For example, considering the sources in the 1 mJy sample over the
interval $1<z<4$, the sources that repeat at least one time have an
average flux of 1.9~mJy, compared to 1.3~mJy for those that do not.
Inspecting some of the nonrepeating sources, we find that they
too are sources with slowly varying star formation rates, which
are just cresting over the sample threshold.  Similar conclusions
apply at higher flux thresholds as well.

\timedistfig

We can also follow the star formation in individual galaxies over
cosmic time.  While there is some rapid variation in the star
formation rates of these galaxies, the amount of such ``noise'' is
typically less than a factor of 2, and the variation in star formation
rate is typically dominated by slow evolution scales of about a Gyr.
(c.f.\ \citealt{weinberg00}).  Our overall conclusion is that our
simulated sub-mm sources are forming large numbers of stars in a
fairly steady way at high redshifts.

Because they are so massive, the sub-mm galaxies in our
simulation are strongly clustered.  Over the redshift range $1<z<4$,
their spatial correlation length is about 4--$5 h^{-1} \Mpc$, with a
correlation power-law exponent $\gamma \approx 2.0$.  For $1<z<3$,
the correlation amplitude is enhanced relative to the average galaxy resolved
in our simulation by a factor of $\sim 1.7$.  However, this strong
clustering will be difficult to detect observationally, because these
sources are spread out over a wide range of redshifts.  Direct
measurement of the redshift-space correlation function requires an
accurate redshift survey of the sub-mm sources, something that will not be
achieved for some time.

We can predict the angular clustering of sub-mm sources using the
simulated maps described at the beginning of \S\ref{sec.observables}.
The angular correlation function of sources with fluxes
above 1~mJy is approximated by
$w(\theta) \approx 0.02 (\theta/ \amin)^{-0.9}$ over a range 0.04 to
10 $\amin$.  This angular clustering is about a factor 3 lower than
that of the Lyman break galaxies \citep{giavalisco98}, which is
accurately reproduced in our simulations \citep{katz99}.  The reason
for the weaker angular clustering of the sub-mm sources is simply that
the stronger three-dimensional clustering is offset by a larger
redshift depth.  To quantify the angular clustering of the 1~mJy
sources to 10\%, a survey area of about 5 square degrees will be
required.  This is a much larger survey than available at present, but
it could be easily achieved with the LMT.

The merger trees for our sub-mm sample allow us to follow the sub-mm
sources in the L50/144 simulation to the present day.  We define a sub-mm
descendant here as a galaxy whose merger tree contains a sub-mm source
in the redshift range $1<z<4$.  Of course, these galaxies often
contain other material as well.  In some cases the particles from a
sub-mm source form only a small part of the mass of a central cD-like
cluster galaxy.

Naturally, we find that the sub-mm descendants are quite massive,
since the sub-mm sources were already quite massive.
We can examine the stellar population of these descendants as well.
The simulation records each star formation event as it takes place, so
we can find the star formation history of any given galaxy.
Figure~\ref{fig.age_z0} plots the median star formation time for each
galaxy in the simulation, versus the total baryonic mass of the
galaxy.  The sub-mm descendants are shown as the more prominent
points.  While the sub-mm descendants are usually massive, there are
some that are among the less massive galaxies.  However, they are
almost universally among the oldest galaxies in the simulation, in
terms of their stellar population.  The vertical lines show the time
range in which the middle 80\% of the stars formed, for a few randomly
chosen galaxies.  One can see that the star formation is quite
extended, agreeing with our depiction of steady star formation.  In
the real universe, the oldest galaxies are giant cluster ellipticals, which
seem to have formed most of their stars before $z \gtrsim 2$
\citep{bower92,stanford95}.
Few galaxies in our simulation have a median star formation redshift
as high as $z = 2$, 
but we only have two galaxy clusters in our simulation and they are
only the size of the Virgo cluster.  Spheroidal populations, i.e.
ellipticals
and bulges, may have a smaller median age \citep{zepf97, fontana99}.
Qualitatively, however, the large ages shown in
the figure suggest an identification with elliptical galaxies.


\agefig

The descendants of the sub-mm sources are also highly clustered.  One
way to measure this is by the circular velocities of their surrounding
environments.  We measure the circular velocities of the virialized
dark matter halos in the same way as discussed above.  Then, we define
``cluster'' and ``group'' halos by the requirements $V_c > 700 \kms$
and $350 \kms < V_c < 700 \kms$ respectively.  With these definitions,
79\% of all galaxies in our L50/144 simulation reside in the field, 18\%
in groups, and 3.2\% in clusters at $z=0$.  In contrast, 57\% of our sub-mm
source descendants reside in the field, 35\% in groups, and 6.6\% in
clusters.  Another way to show the strong clustering is to examine the
fraction of galaxies that are sub-mm descendants as a function of
local galaxy number density.  We find a trend quite similar to the
well-known density-morphology relation (e.g., \citealt{postman84}),
but even more extreme, with almost no descendants below number
densities of $0.1 h^3 \Mpc^{-3}$.

It has been suggested that the sub-mm sources are massive ellipticals
in the process of formation.  One argument for this is that elliptical
galaxies have old stellar populations, as shown (at least in clusters)
by their tight color-magnitude relation \citep{bower92} and the
passive evolution of their colors \citep{stanford95}.  This suggests,
although it does not prove, that the galaxies themselves were formed at
high redshift.  Another argument is the common hypothesis that
ellipticals were formed in mergers, coupled with the observation that
ULIRGs at low redshift are typically mergers in progress.  If the
high-redshift sub-mm sources are similar to the low-redshift ULIRGs,
this supports the notion that they mark the birth of ellipticals.

In the L50/144 simulation, our spatial resolution is too low to resolve the
morphology of our galaxies and distinguish spirals from ellipticals.
However, several features of the sub-mm source descendants---high mass, old
stellar populations, and strong clustering---do indeed suggest that
they correspond to massive ellipticals in the process of formation.
Still, we emphasize that in our simulations the sources usually do not
owe their high luminosities to very short bursts.  Our sub-mm
galaxies instead form in a more gradual manner over periods of $\tsim
0.1$--$1 \Gyr$.

\section{DISCUSSION}
\label{sec.discussion}

The results presented in this paper amount to a physically motivated
model for the population of sub-mm galaxies detected in the SCUBA surveys.
The underlying basis of this model is the $\Lambda$-dominated cold dark 
matter cosmological scenario, coupled to our numerical methods for following
gravitational evolution, gas dynamics and cooling, and star formation.
The essential features of the star formation algorithm are a gas consumption
timescale that decreases steadily with increasing gas density, feedback 
effects that are relatively modest, at least in the massive galaxies
that correspond to sub-mm sources, and an IMF similar to that inferred
in the solar neighborhood \citep{miller79}.
Given our numerical simulation of the population of star-forming
galaxies, the main uncertainty in our predictions comes from the
choice of FIR spectral energy distribution.  Matching the observed
sub-mm counts requires values of the SED parameters $T$ and $\beta$
that are fairly close to those inferred for IR-luminous galaxies
in the local universe by \citet{dunne00}, but these values are lower
than those used in most models of the SCUBA population.
Our default model has $T=35\K$, $\beta=1.0$, but the values of
$T$ and $\beta$ have nearly degenerate effects, scatter in the SEDs
would yield similar counts for higher $T$ and/or $\beta$ values,
and the overall factor for converting star formation rate to bolometric
luminosity is itself uncertain at the factor of two level.

The essential features of this model are that the sub-mm population
is broadly distributed in redshift, with a median $z \approx 2-2.5$,
and consists mainly of massive galaxies forming stars fairly steadily
over timescales $\sim 10^8-10^9$ years at rates of $\sim 100\sfrunit$.
The descendants of these sub-mm sources are even more massive galaxies,
with old stellar populations, found primarily in dense environments.
At a qualitative level, these properties support the identification
of sub-mm sources with the progenitors of luminous early-type galaxies.

The steady star formation and correspondingly moderate star formation
rates are what distinguish this model of the sub-mm population from
some of the alternative pictures proposed in the literature
(discussed further below).  While the typical formation timescales
are still a factor of a few shorter than the cosmic time at the same
redshift, they are much larger than, e.g., the timescale of 
$\sim 2\times 10^6$ yrs suggested for FIR emission by \citet{thronson86}.
Typical star formation rates of $\sim 100\sfrunit$ require, in turn,
relatively low values of $T$ and/or $\beta$ to yield sufficient flux
at sub-mm wavelengths.

The principal successes of this model are its ability to reproduce the
bright ($>1\mJy$) sub-mm counts and its consistency with existing
constraints on the source redshift distribution (Figure~\ref{fig.zdist}).
Unfortunately, the first cannot be counted as a major success because
of the large observational uncertainties and the substantial freedom
to change the predicted fluxes by reasonable alternative SED choices.
We did not make any adjustments to match the redshift distribution,
so success there is reassuring, though the model does predict a 
potentially significant excess of low-redshift sources.  The model
also has difficulty matching source counts above $10\mJy$, though
these could be affected observationally by contributions from AGN
and by lack of large-scale power in our finite simulation 
volume.

Another, potentially more serious failing of this model is its prediction,
once we correct for numerical resolution effects, of too many faint 
sources and an excessively high 850~\micron\ background
(see Figure~\ref{fig.medrescounts}).  We will save detailed discussion
of this issue for a future paper \citep{fardal01}, but a few comments
are in order here.  
Part of this problem may simply be due to the small volume 
of the L11/64 simulation we 
used to constrain the properties of smaller objects at low redshifts.
Overproduction of faint counts and the sub-mm
background, and the possible excess of low redshift sources mentioned
above, could be connected to other possible shortcomings of the
simulated galaxy population that we have noted elsewhere
(e.g., \citealt{katz96,weinberg99,aguirre01}):
an excess of low redshift star formation and an overly high fraction of
baryons converted into stars.  
\cite{balogh01} have argued that the
latter problem is generic to simulations of this sort and can only
be solved by appealing to much more vigorous feedback effects from star
formation (see \citealt{cole00} for discussion of similar problems
in semi-analytic models).  While we agree that feedback could be
a solution, there are many other possibilities, including: observational
errors in the stellar density (see, e.g., the discussions of luminosity
function discrepancies by \citealt{cole01}, \citealt{blanton01},
and \citealt{wright01}); numerical errors causing the simulations
to systematically overestimate gas cooling; changes to the cosmological
model such as reduced initial power on galactic scales or lower
baryon density; or an initial mass function (at least in some galaxies)
that contains a large mass fraction of brown dwarfs 
or differs systematically from the solar neighborhood IMF in some other way.
We note also that if feedback {\em alone} is to 
reconcile the simulations with the sub-mm
source counts at $\tsim 1\mJy$, then it must be effective in galaxies
with baryon masses $\tsim 5\times 10^{10} M_\odot$ (see 
Figure~\ref{fig.massfunc}), not just in low mass systems.

An obvious risk is that any change to our model that reduces the faint
source counts and sub-mm background will also spoil the agreement with
bright source counts.  A luminosity-dependent SED is possible, and
perhaps even probable.  A luminosity-dependent IMF could similarly
improve the agreement.  Both, however, represent unattractively {\it
ad hoc} solutions at present.
Another generic way to make the predicted source counts flatter
would be to make star formation episodic, so that more of the
bright sources correspond to more common, low mass objects caught
during a burst of star formation.  In particular, it is possible
that our simulations underestimate the importance of merger-induced
starbursts because of their limited resolution.
In their simulations of galaxy mergers, \citet{mihos95} and 
\citet{mihos96} found bursts of star formation enhanced by a factor
10--100 over the star formation in isolated galaxies.  The resolution
in our cosmological simulation may simply be too poor to obtain such large
bursts, or the large particle-induced noise may prevent the buildup of
large gas
reservoirs.  Whereas each galaxy in the Mihos \& Hernquist simulations
was represented by about $3\times 10^4$ gas particles, our galaxies
often contain only a few hundred, and they do not resolve galactic disks.
Observationally, we know that
low-redshift mergers produce enormous bursts of star formation
\citep{sanders96}.  

On the other hand, there are reasons to suspect 
that bursts and mergers do not play a dominant role in the
global history of star formation, even if they are important in some objects.
From this same L50/144 simulation,
\citet{murali01} conclude that smooth accretion dominates over
merging by at least 3:1 in mass during the assembly of galaxies.  While
the mass limit of the simulation is quite high, making it impossible
to distinguish between low-mass mergers and accretion, they argue
from the mass spectrum of the merging objects that true accretion is
still dominant.  The effect of individual mergers may also be less dramatic at
high redshifts than at the present day.  Cosmological timescales at
high $z$ are small (e.g., the time from $z=3$ to $z=2$ is only
1.1~Gyr), and every merger takes some finite amount of time $\sim
0.1$--1~Gyr.  Moreover, the merger rate of galaxies is larger at high
redshift.  Thus the bursts induced by different merger events may
overlap for long stretches at high redshifts.  From studying the star
formation histories alone, it may be unclear whether the star
formation is steady because mergers are unimportant, or because they
are ubiquitous.
In any case, higher-resolution simulations and detailed observations
of high-redshift galaxies should clarify the role of mergers in
producing high-luminosity sources.

There have been a number of models for the sub-millimeter sources in
the literature, mostly phenomenological in nature.  For example,
\citet{blain99model} constructed sets of models based on pure
luminosity evolution of the IRAS low-redshift luminosity function.
They constrained the average SED and the low-redshift evolution by
comparing the 60 \micron\ and 175 \micron\ counts from low-redshift
sources.  They then found several fitting functions for the typical
luminosity, with ``anvil'' and ``peak'' shapes, that adequately
represented the sub-mm counts.  In a follow-up paper,
\citet{blain99semimodel} used a Press-Schechter formalism for merging
objects to describe the sub-mm sources.  To enable a fit to the
counts, similarly peaked fitting functions were used for the amount of
stars formed in an average merger.  Unfortunately, the power spectra
and dark halo masses used in this formalism appear somewhat
disconnected from those expected within the CDM model, and the
physical nature of the sub-mm sources in this model is somewhat
obscure.  Both of these papers put the sub-mm sources at somewhat
higher redshifts (median $z$ of 2.5--4.5) than in our work, and derive
much higher star formation rates (both in individual objects and in
total).  This is partly because of the different SED, and partly from
the smaller assumed energy output from star formation.

Recently, \citet{shu01} have constructed a simpler analytic model for
the sub-mm sources, in a sense performing the usual semi-analytical
calculation backwards.  They start from the observed distributions of
the sizes and star formation rates of Lyman break galaxies, and use
the Schmidt law for galactic disks to derive the gas masses, circular
velocities, and star formation timescales.  Within this model, sub-mm
sources are simply defined to be Lyman break galaxies with 
${\it SFR} > 1000 \sfrunit$ (corresponding to the bright end of the
observed SCUBA sources), and are assumed to have a top hat redshift
distribution from $2.5 < z < 3.5$.  The brightest sub-mm galaxies in
their model are quite massive and highly correlated ($r_0 \approx 7 \,
h^{-1} \Mpc$), with even more highly clustered descendants
corresponding to giant cluster ellipticals.  The source timescales are
a few tenths of a Hubble time, probably consistent with our results.
Despite the crude approach, the basic picture they derive is fairly
consistent with that drawn in this paper. 

Taken at face value, our model makes several testable predictions.
Probably the most robust --- because it is directly connected to 
the long timescale and moderate rate of star formation --- is that
the FIR SED parameters of SCUBA sources should be ones that produce
a relatively large 850~\micron\ flux for a given star formation rate.
Rest-frame UV and optical observations of sub-mm sources may also
yield constraints on the star formation rates and timescales in these systems.
Improved constraints on the redshift distribution of sub-mm sources
can also test the model more stringently and give clearer guidance
to the origin of possible discrepancies.  Finally, our model predicts
that the sub-mm galaxies are strongly clustered: their redshift-space
clustering should exceed that of typical Lyman-break galaxies, though
their angular clustering will be lower because of the large redshift range.
Testing these predictions will require substantial improvements
in the multi-wavelength and spectroscopic identification of sub-mm
sources.  These improvements will not come easily, but we have gone
from knowing essentially nothing about the sub-mm source population
to knowing quite a lot in the space of a few years, and we can expect
similar improvements in the years to come.

We thank Simon Lilly, Amy Barger, David Hughes, Stephen Eales, Andrew
Blain, and Scott Chapman for helpful discussions.  
We also thank Jeff Gardner for the use of his program SO.
This work was supported
by NASA Astrophysical Theory Grants NAG5-3922, NAG5-3820, NAG-4064,
and NAG5-3111.
By NASA Long-Term Space Astrophysics Grant NAG5-3525, and by the NSF under
grants ASC93-18185, ACI96-19019, and AST-9802568.
The simulations were performed at the San Diego Supercomputer Center,
NCSA, and the NASA/Ames Research Center.


\begin{thebibliography}{}

\bibitem[Adelberger \& Steidel(2000)]{adelberger00} Adelberger, K.\ 
L.\ \& Steidel, C.\ C.\ 2000, \apj, 544, 218 

\bibitem[Aguirre et al.(2001)]{aguirre01}
Aguirre, A., Hernquist, L., Schaye, J., Weinberg, D. H., 
Katz, N., \& Gardner, J. 2001, \apj, submitted, astro-ph/0105065

\bibitem[Balogh et al.(2001)]{balogh01}
Balogh, M., Pearce, F., Bower, R., \& Kay, S. 2001, \mnras, in press,
astro-ph/0104041

\bibitem[Barger et al.(1998)]{barger98} Barger, A.\ J., Cowie, 
L.\ L., Sanders, D.\ B., Fulton, E., Taniguchi, Y., Sato, Y., Kawara, K.\ 
\& Okuda, H.\ 1998, \nat, 394, 248 

\bibitem[Barger, Cowie \& Sanders(1999)]{barger99} Barger, A. 
J., Cowie, L. L. \& Sanders, D. B. 1999, \apjl, 518, L5 

\bibitem[Barger, Cowie \& Richards(2000)]{barger00} Barger, 
A. J., Cowie, L. L. \& Richards, E. A. 2000, \aj, 119, 2092

\bibitem[Barger, Cowie, Mushotzky, \& Richards(2001)]{barger01} 
Barger, A.\ J., Cowie, L.\ L., Mushotzky, R.\ F., \& Richards, E.\ A.\ 
2001, \aj, 121, 662 

\bibitem[Blain(1999)]{blain99temp} 
Blain, A.\ W.\ 1999, \mnras, 309, 955 

\bibitem[Blain et al.(1999a)]{blain99review} Blain, 
A.\ W., Smail, I., Ivison, R.\ J.\ \& Kneib, J.-P.\ 1999, ASP Conf.\ 
Ser.\ 193: The Hy-Redshift Universe: Galaxy Formation and Evolution at High 
Redshift, 425 

\bibitem[Blain et al.(1999b)]{blain99model} Blain, 
A. W., Smail, I., Ivison, R. J. \& Kneib, J.-P.\ 1999, \mnras, 302, 632 

\bibitem[Blain et al.(1999c)]{blain99semimodel} Blain, A.\ W., Jameson, 
A., Smail, I., Longair, M.\ S., Kneib, J.-P., \& Ivison, R.\ J.\ 1999, 
\mnras, 309, 715 

\bibitem[Blain, Ivison, Kneib \& Smail(1999)]{blain99obs2} 
Blain, A.\ W., Ivison, R.\ J., Kneib, J.-P.\ \& Smail, I.\ 
1999, ASP Conf.\ Ser.\ 193: The Hy-Redshift Universe: 
Galaxy Formation and Evolution at High Redshift, 246 

\bibitem[Blain(2000)]{blain00} 
Blain, A. W.  2000, astro-ph/0011479, from Deep Fields meeting
[REF TO BE FIXED]

\bibitem[Blanton et al.(2001)]{blanton01}
Blanton, M.\ R.\ et al.\ 2001, \aj, 121, 2358

\bibitem[Borys et al.(2001)]{borys01} 
Borys, C., Halpern, M., Chapman, S., \& Scott, D. 
  2001, in Deep Millimeter Surveys, 
  ed.\ J. D. Lowenthal, D. Hughes, World Scientific (NY, Tokyo), 
  astro-ph/0009143

\bibitem[Bower, Lucey \& Ellis(1992)]{bower92} Bower, R.\ G., 
Lucey, J.\ R.\ \& Ellis, R.\ S.\ 1992, \mnras, 254, 601 

\bibitem[Bullock et al.(2001)]{bullock01}
Bullock, J. et al., 2001, in preparation

\bibitem[Burles \& Tytler(1997)]{burles97}
Burles, S., \& Tytler, D. 1997, \aj, 114, 1330

\bibitem[Burles \& Tytler(1998)]{burles98}
Burles, S., \& Tytler, D. 1998, \apj, 499, 699

\bibitem[Chapman et al.(2000a)]{chapman00survey}
Chapman, S. C., Scott D., Borys C., Fahlman. G. G, 2000a, astro-ph/0009067

\bibitem[Chapman et al.(2000)]{chapman00radiosel}
Chapman, S. C., Richards, E., Lewis, G. F., Wilson, G., 
\& Barger, A.  2000b, astro-ph/0011066

\bibitem[Cole et al.(2001)]{cole01}
Cole, S., et al. 2001, \mnras, in press, astro-ph/0012429

\bibitem[Cole et al.(2000)]{cole00}
Cole, S., Lacey, C.\ G., Baugh, C.\ M., \& Frenk, C.\ S.\ 2000, \mnras, 319, 168

\bibitem[Croft et al.(2000)]{croft00}
Croft, A. C., di Matteo, T., Dav\'{e}, R., Hernquist, L., Katz, N., 
Fardal, M. A., Weinberg, D. H. 2000, ApJ, in press [astro-ph/0010345]

\bibitem[Dav\'e, Dubinski, \& Hernquist(1997)]{dave97code}
Dav\'e, R., Dubinski, J., \& Hernquist, L. 1997,
New Astron, 2, 227

\bibitem[Dav\'e et al.(1999a)]{dave99}
Dav\'e, R., Hernquist, L., Katz, N., \& Weinberg, D. H. 1999a, \apj, 511, 521

\bibitem[Dav\'e et al.(1999b)]{davelbg99}
Dav\'e, R., Gardner, J.P., Hernquist, L., Katz, N., \& Weinberg, D. H. 
1999b, in the Proceedings of Rencontres Internationales de l'IGRAP,
Clustering at High Redshift, Marseille 1999 [astro-ph/9910220]


\bibitem[Doane \& Mathews(1993)]{doane93} Doane, J.\ S.\ \& 
Mathews, W.\ G.\ 1993, \apj, 419, 573 

\bibitem[Dunlop(2001)]{dunlop01} Dunlop, J. S.
  2001, in Deep Millimeter Surveys, 
  ed.\ J. D. Lowenthal, D. Hughes, World Scientific (NY, Tokyo), 
  astro-ph/0011077

\bibitem[Dunne et al.(2000)]{dunne00} Dunne, L., Eales, S., 
Edmunds, M., Ivison, R., Alexander, P. \& Clements, D. L. 2000, \mnras, 
315, 115 

\bibitem[Eales et al.(2000)]{eales00} Eales, S., Lilly, S., 
Webb, T., Dunne, L., Gear, W., Clements, D.\ \& Yun, M.\ 2000, \aj, 120, 
2244 

\bibitem[Fardal et al.(2001)]{fardal01} 
Fardal, M.A. et al., 2001, in preparation

\bibitem[Fixsen et al.(1998)]{fixsen98} Fixsen, D. J., Dwek, 
E., Mather, J. C., Bennett, C. L. \& Shafer, R. A. 1998, \apj, 508, 
123 

\bibitem[Fontana et al.(1999)]{fontana99}
Fontana, A., Menci, N., D'Odorico, S., Giallongo, E., Poli, F., Cristiani, S.,
Moorwood, A., \& Saracco, P. 1999, \mnras, 310, 27

\bibitem[Gardner et al.(1997a)]{gard97a}
Gardner, J.P., Katz, N., Hernquist, L. \& Weinberg, D.H.\ 1997a, \apj,
484, 31

\bibitem[Gardner et al.(1997b)]{gard97b}
Gardner, J.P., Katz, N., Weinberg, D.H. \& Hernquist, L.\ 1997b, \apj,
486, 42

\bibitem[Gardner et al.(2001)]{gard01}
Gardner, J.P., Katz, N., Hernquist, L. \& Weinberg, D.H.\ 2001, \apj,
in press [astro-ph/9911343]


\bibitem[Giavalisco et al.(1998)]{giavalisco98} Giavalisco, M., 
Steidel, C.\ C., Adelberger, K.\ L., Dickinson, M.\ E., Pettini, M.\ \& 
Kellogg, M.\ 1998, \apj, 503, 543 

\bibitem[Guiderdoni et al.(1998)]{guiderdoni98} 
Guiderdoni, B., Hivon, E., Bouchet, F.\ R.\ \& Maffei, B.\ 1998, 
\mnras, 295, 877 

\bibitem[Hauser et al.(1998)]{hauser98} Hauser, M.\ G.\ et al.\ 
1998, \apj, 508, 25 

\bibitem[Hernquist \& Katz(1989)]{hk89}
Hernquist, L. \& Katz, N.\ 1989, \apjs, 70, 419

\bibitem[Hernquist et al.(1996)]{hernquist96}  
Hernquist L., Katz, N., Weinberg, D.H., \&
Miralda-Escud\'e, J. 1996, \apj, 457, L5

\bibitem[Hu \& Sugiyama(1996)]{hu96}
Hu, W. \& Sugiyama, N. 1996, \apj, 471, 542

\bibitem[Hughes et al.(1998)]{hughes98} Hughes, D.\ H.\ et al.
1998, \nat, 394, 241 

\bibitem[Hughes \& Gazta\~{n}aga(2000)]{hughes00}
Hughes, D. H., \& Gazta\~{n}aga, E.
2000, Star formation on Large-scales to Small-scales, 
ed.\ F. Favata, A. A. Kaas \& A. Wilson, ESA

\bibitem[Ivison et al.(1998)]{ivison98} Ivison, R.\ J., Smail, 
I., Le Borgne, J.-F., Blain, A.\ W., Kneib, J.-P., Bezecourt, J., Kerr, 
T.\ H.\ \& Davies, J.\ K.\ 1998, \mnras, 298, 583 

\bibitem[Katz, Weinberg, \& Hernquist(1996)]{katz96} 
Katz, N., Weinberg D.H., \& Hernquist, L. 1996, \apjs, 105, 19 

\bibitem[Katz, Hernquist, \& Weinberg(1999)]{katz99}
Katz, N., Hernquist, L., \& Weinberg, D. H. 1999, \apj, 523, 463

\bibitem[Kitayama \& Suto(1996)]{kitayama96}
Kitayama, T., \& Suto, Y. 1996, \apj, 469, 480

\bibitem[Kroupa(2000)]{kroupa00}
Kroupa, P. 2000, Dynamics of Star Clusters and the Milky Way, 
ed.\ S. Deiters, R. Spurzem, et al.

\bibitem[Leitherer et al.(1999)]{leitherer99} 
Leitherer, C. , Schaerer, D., Goldader, J. D., Gonz\'{a}lez Delgado, R. M., 
Robert, C., Kune, D. F., de Mello, D. F., Devost, D., \&
Heckman, T. M. 1999, \apjs, 123, 3 

\bibitem[Lilly et al.(1999)]{lilly99} Lilly, S.\ J., Eales, 
S.\ A., Gear, W.\ K.\ P., Hammer, F., Le F{\`e}vre, O., Crampton, D., 
Bond, J.\ R.\ \& Dunne, L.\ 1999, \apj, 518, 641 

\bibitem[Madau, Pozzetti \& Dickinson(1998)]{madau98} Madau, 
P., Pozzetti, L.\ \& Dickinson, M.\ 1998, \apj, 498, 106 

\bibitem[Hernquist \& Mihos(1995)]{mihos95} Hernquist, L.
\& Mihos, J. C. 1995, \apj, 448, 41 

\bibitem[Mihos \& Hernquist(1994a)]{mihos94a} Mihos, J. C. 
\& Hernquist, L. 1994a, \apj, 431, L9 

\bibitem[Mihos \& Hernquist(1994b)]{mihos94b} Mihos, J. C. 
\& Hernquist, L. 1994b, \apj, 425, L13 

\bibitem[Mihos \& Hernquist(1996)]{mihos96} Mihos, J. C. 
\& Hernquist, L. 1996, \apj, 464, 641 

\bibitem[Miller \& Scalo(1979)]{miller79}
Miller, G. E., \& Scalo, J. M.  1979, ApJS, 41, 513

\bibitem[Murali et al.(2001)]{murali01} 
Murali, C., Katz, N., et al. 2001, in preparation

\bibitem[Postman \& Geller(1984)]{postman84} Postman, M.,
\& Geller, M. J. 1984, ApJ, 281, 95

\bibitem[Rauch et al.(1997)]{rauch97} Rauch, M., Miralda-Escud\'e, J., 
Sargent, W.L.W., Barlow, T.A.,Hernquist, L., Weinberg D.H., Katz, N., 
Cen, R., Ostriker, J.P. 1997, \apj, 489, 7

\bibitem[Sanders \& Mirabel(1996)]{sanders96} Sanders, D.\ B.\ 
and Mirabel, I.\ F.\ 1996, \araa, 34, 749 

\bibitem[Saunders et al.(1990)]{saunders90} Saunders, W., 
Rowan-Robinson, M., Lawrence, A., Efstathiou, G., Kaiser, N., Ellis, R.
S. \& Frenk, C. S. 1990, \mnras, 242, 318 

\bibitem[Schechter(1976)]{schechter76}
Schechter, P. 1976, \apj, 203, 297

\bibitem[Seljak \& Zaldarriaga(1996)]{seljak96}
Seljak, U., \& Zaldarriaga, M. 1996, \apj, 469, 437

\bibitem[Shu, Mao, \& Mo(2001)]{shu01}
Shu, C., Mao, S., \& Mo, H. J. 2001, astro-ph/0102436

\bibitem[Smail, Ivison \& Blain(1997)]{smail97} Smail, I., 
Ivison, R.\ J.\ \& Blain, A.\ W.\ 1997, \apjl, 490, L5 

\bibitem[Smail et al.(2000)]{smail00zdist} Smail, I., Ivison, R.\ 
J., Owen, F.\ N., Blain, A.\ W.\ \& Kneib, J.-P.\ 2000, \apj, 528, 612 

\bibitem[Stanford, Eisenhardt \& Dickinson(1995)]{stanford95} 
Stanford, S.\ A., Eisenhardt, P.\ R.\ M.\ \& Dickinson, M.\ 1995, \apj, 
450, 512 

\bibitem[Sullivan et al.(2000)]{sullivan00} Sullivan, M., Treyer, 
M.\ A., Ellis, R.\ S., Bridges, T.\ J., Milliard, B., \& Donas, J.\ 2000, 
\mnras, 312, 442 

\bibitem[Thronson \& Telesco(1986)]{thronson86} Thronson, H.\ A.\ 
\& Telesco, C.\ M.\ 1986, \apj, 311, 98 

\bibitem[Weinberg et al.(1999)]{weinberg99} Weinberg, D. H., 
Dav{\'e}, R., Gardner, J. P., Hernquist, L. \& Katz, N. 1999, ASP 
Conf.\ Ser.\ 191: Photometric Redshifts and the Detection of High Redshift 
Galaxies, 341 

\bibitem[Weinberg, Hernquist, \& Katz(2000)]{weinberg00} Weinberg, D. H.,
Hernquist, L., \& Katz, N.  2000, astro-ph/0005340 [REF TO BE FIXED]

\bibitem[White, Efstathiou, \& Frenk(1993)]{white93}
White, S. D. M., Efstathiou, G. P., \& Frenk, C. S. 1993, \mnras, 262, 1023

\bibitem[Wright(2001)]{wright01}
Wright, E. L. 2001, \apj, submitted, astro-ph/0102053

\bibitem[Yun, Reddy, \& Condon(2001)]{yun01} Yun, M.\ S., 
Reddy, N.\ A., \& Condon, J.\ J.\ 2001, \apj, 554, 803 

\bibitem[Zaldarriaga, Seljak, \& Bertschinger(1998)]{zaldarriaga98}
Zaldarriaga, M., Seljak, U., \& Bertschinger, E. 1998, \apj, 494, 491

\bibitem[Zepf(1997)]{zepf97}
Zepf, S. 1997, Nature, 390, 377

\end{thebibliography}
\end{document}